\documentclass[reprint,aps,amsmath,amssymb,superscriptaddress,pra]{revtex4-2}
\usepackage{graphicx}
\usepackage{dcolumn}
\usepackage{bm}
\usepackage{color}
\usepackage{amsmath}
\usepackage{booktabs} 
\usepackage{array} 
\usepackage{makecell}
\usepackage{colortbl}
\usepackage{xcolor} 
\usepackage[justification=centering,singlelinecheck=false]{caption}
\usepackage{braket}
\usepackage{subcaption}
\usepackage{caption}
\usepackage[english]{babel}

\newcommand{\vrr}{\mathbf{r}}
\newcommand{\vs}{\mathbf{s}}
\newcommand{\vd}{\mathbf{d}}
\newcommand{\vRR}{\mathbf{R}}
\newcommand{\vG}{\mathbf{G}}
\newcommand{\vL}{\mathbf{L}}
\newcommand{\vT}{\mathbf{T}}
\newcommand{\vM}{\mathbf{M}}
\newcommand{\vm}{\mathbf{m}}
\newcommand{\vn}{\mathbf{n}}
\newcommand{\vk}{\mathbf{k}}
\newcommand{\ve}{\mathbf{e}}
\newcommand{\vu}{\mathbf{u}}
\newcommand{\vP}{\mathbf{P}}
\newcommand{\vx}{\mathbf{x}}
\newcommand{\At}{\widetilde{A}}
\newcommand{\Tr}{\mathrm{Tr}}

\begin{document}
\title{Explicitly Correlated Gaussian Basis Approach to Periodic
Systems}

\author{K\'alm\'an Varga}
\email{kalman.varga@vanderbilt.edu (corresponding author)}
\affiliation{Department of Physics and Astronomy, Vanderbilt
University, Nashville, Tennessee, 37235, USA}
\begin{abstract}
Closed-form expressions for all matrix elements required for
variational calculation of the electronic structure of periodic solids
have been derived using a basis of explicitly correlated Gaussians
(ECGs). Periodic basis functions are constructed by summing shifted
correlated Gaussians over all composite lattice translations, where a
generalized unfolding theorem reduces the resulting double lattice sum
to a single sum through a unified computational framework for overlap,
kinetic energy, and Coulomb potential operators. 
The formalism has been validated through
application to an infinite one-dimensional hydrogen chain, where the
ground-state energy per atom computed in the thermodynamic limit is
shown to agree with finite-chain results extrapolated by
other many-body methods.
\end{abstract}

\maketitle

\section{Introduction}
Few-body methods based on Explicitly Correlated Gaussians (ECGs) have
become 
indispensable tools for high-accuracy calculations in atomic and
molecular 
physics \cite{boys60,singer60,kolos63a,drake70a,drake91a,yan97a,
korobov00a,nakatsuji07a,ryzhikh97b,PhysRevA.84.012509,bubin04a,stanke06a,
cencek95b,1.4794192,1.4826450,1.4834596,cr200419d,1.4897634,1.4873916},
enabling precise characterization of electron correlations
\cite{lin83a}, 
relativistic effects \cite{PhysRevA.84.012509,cencek96a}, molecular
bonding 
\cite{richard94a,strasburger99a,cencek00a}, and nuclear quantum
dynamics 
\cite{stanke07d,bubin07b,pachucki09a,holka11a}. The power of this
approach is 
exemplified by the 1~MHz-level agreement between theory 
\cite{PhysRevLett.122.103003} and experiment
\cite{PhysRevLett.122.103002} for 
the dissociation energy of H$_2$, enabling tests of fundamental
constants and 
guiding the development of efficient approximate methods.

ECGs have been in use since 1960 \cite{boys60,singer60} and owe their 
popularity
\cite{PhysRevA.102.062825,PhysRevLett.118.043001,PhysRevA.100.032504,
1.4826450,B211193D,4890373,varga_1998,3491029,1.4834596,cr200419d,
PhysRevLett.113.213201,1.4897634,RevModPhys.85.693,PhysRevA.92.013608,
PhysRevA.89.052501,PhysRevA.84.012509,PhysRevLett.113.073004,
PhysRevA.89.032510,PhysRevA.92.012513,1.4873916,4731696,PhysRevA.80.062510,
ZaklamaTimothy2020MEoO,ctx19577663200003276,KEDZIORSKI2020137476,
ctx19578016840003276,MATYUSEdit2013OtCo,doi:10.1063/5.0051237,
PhysRevA.102.052806,PhysRevA.102.022803,PhysRevE.101.023313,
NASIRI2020143,PhysRevA.100.042503,Stanke_2019,M_ller_2019,PhysRevA.99.012504,
doi:10.1063/1.5050462,doi:10.1063/1.5009465,ADAMOWICZ201787,
PhysRevA.95.062509,Fedorov2016,doi:10.1063/1.4948708,BUBIN2016122,
PhysRevA.92.062501,PhysRevA.89.012506,PhysRevLett.111.193401,
PhysRevA.87.054501,PhysRevD.103.074503,
doi:10.1021/acs.jctc.5c00970,doi:10.1021/acsphyschemau.5c00055}
to the fact that their inter-particle quadratic form yields
analytically simple 
Hamiltonian matrix elements whose algebraic complexity is independent
of 
particle number. Matrix elements can be extended to arbitrary angular
momentum 
\cite{doi:10.1063/1.4948708,Stanke_2019,4731696,sharkey_pra_80_062510_2009,
Varga1998,4890373} and the Gaussian parameters are optimized
variationally 
\cite{kozlowski92b,suzuki98a,komasa95,PhysRevA.99.012504,bubin10a,tung11a,
sharkey11c,bubin2008}, with extensions to complex parameters 
\cite{bubin06b,PhysRevA.95.062509,BUBIN2016122} and periodic boundary 
conditions \cite{PhysRevA.87.063609} also explored.

The scope of ECG applications, reviewed comprehensively in 
Refs.~\cite{cr200419d,B211193D,RevModPhys.85.693}, spans spherical
($L=0$) 
systems
\cite{suzuki98a,kozlowski91a,kozlowski92a,cencek93a,cencek95a,varga95,
bubin2008,PhysRevD.103.074503,PhysRevA.87.063609}---including Efimov
physics 
\cite{PhysRevLett.113.213201}, hyperfine splitting 
\cite{PhysRevLett.111.243001,PhysRevA.89.032510}, QED corrections 
\cite{PhysRevA.92.012513}, and cold Fermi gases 
\cite{PhysRevA.92.013608}---as well as large few-body systems such as 
H$_3^+$ \cite{ctx19577663200003276}, Be \cite{PhysRevA.100.032504}, B 
\cite{PhysRevLett.118.043001}, and C$^+$ \cite{PhysRevA.102.062825},
the last 
requiring up to 16000 basis functions. ECGs have also been applied to
nuclear 
cluster models \cite{suzuki2008,PhysRevC.89.064303,
PhysRevC.100.024334,10.1093/ptep/pts015,AoyamaS2012FSwa} and extended to
scattering 
problems via the confined variational method 
\cite{PhysRevA.103.052803,PhysRevLett.101.123201,PhysRevA.103.022817,
PhysRevA.103.042814,PhysRevA.101.042705,PhysRevA.100.032701,
PhysRevLett.103.223202,PhysRevA.78.042705}.

For nonspherical ($L>0$) systems, two main strategies exist: shifted
Gaussian 
centers, which keep matrix elements simple but require angular momentum 
projection or explicit construction 
\cite{PhysRevA.102.022803,doi:10.1063/1.5050462,suzuki98a,4890373,
doi:10.1063/1.5009465,doi:10.1080/00268976.2013.783938,
PhysRevA.99.052512,PhysRevA.102.052806}, and multiplication by
polynomials of 
interparticle coordinates. The latter includes $L$-specific
formulations 
\cite{PhysRevA.80.062510,sharkey11b,sharkey11c,sharkey11d,
sharkey_jcp_132_184106_2010}, the global-vector representation 
\cite{suzuki98a,varga95,varga_1998,SuzukiY1998Ndoo,suzuki2008,4731696,
MATYUSEdit2013OtCo}, tensor-product extensions from analytic 1D matrix
elements 
\cite{ZaklamaTimothy2020MEoO}, and a fully general formalism for
arbitrary 
products of single-particle coordinates \cite{doi:10.1063/1.4948708}.

In this paper, we derive closed-form expressions for all matrix elements required for a
variational calculation of the electronic structure of a periodic solid
using a basis of shifted correlated Gaussians (SCGs),
\begin{equation}\label{eq:scg_intro}
  \phi_k(\vrr) = \exp\!\bigl[-(\vrr-\vs_k)^T(A_k\otimes I_3)(\vrr-\vs_k)\bigr],
\end{equation}
where $\vrr=(\vrr_1,\dots,\vrr_n)^T\in\mathbb{R}^{3n}$ collects all $n$
electron coordinates, $A_k\in\mathbb{R}^{n\times n}$ is a symmetric
positive-definite correlation matrix whose off-diagonal elements encode
electron–electron correlations explicitly, and $\vs_k\in\mathbb{R}^{3n}$
is a shift vector that centres the Gaussian.  These functions are
periodized by summing over all composite lattice translations,
\begin{equation}\label{eq:periodic_basis_intro}
  \Phi_k(\vrr) = \sum_{\vM\in\mathbb{Z}^{3n}}\phi_k(\vrr-\vT_{\vM}),
\end{equation}
where $\vT_{\vM}$ displaces all electrons simultaneously by integer
multiples of the simulation-cell dimensions (defined precisely in
Sec.~\ref{sec:basis}).  To our knowledge,
the only prior application of ECGs to a periodic system is the work of
Ref.~\cite{PhysRevA.87.063609}, which treats small two-component Fermi
gases in a cubic box with periodic boundary conditions.  That work,
however, uses a short-range contact (delta-function) interaction: the
particles interact only at zero separation, and no long-range Coulomb
potential is present.  The physically relevant and far more challenging
problem — electrons and nuclei interacting through the long-range $1/r$
Coulomb potential under periodic boundary conditions — requires Ewald
summation or an equivalent regularization, and the corresponding
ECG matrix elements have not previously been derived.  
In this paper, we address the long-range Coulomb interaction using
three complementary approaches: Ewald summation, a direct
neutral-shell sum, and a Dirac delta convolution method capable of
simultaneously determining pair-correlation and contact densities.
Within the Ewald framework, divergences in individual terms cancel
exactly, yielding a finite energy expression that is independent of
computational parameters. These three independent methods yield mutually consistent results, 
confirming the robustness of the approach.

A further distinguishing feature of the electronic-structure problem is that the
energy of a periodic solid is not a single number but a function of the
Bloch wave vector $\mathbf{k}_\mathrm{B}$: computing the band structure
$E_n(\mathbf{k}_\mathrm{B})$ requires evaluating the Hamiltonian and
overlap matrix elements at many $\mathbf{k}_\mathrm{B}$ points in the
first Brillouin zone, which we accommodate through the Bloch-phase
generalization of the unfolding theorem.

Periodization introduces a double lattice sum over image cells.  Using a 
generalized unfolding theorem we reduce this double sum
to a single sum for any lattice-periodic operator, with a Bloch-phase
extension for $k$-point sampling.  After unfolding, every matrix element
factors into a Gaussian prefactor times an image sum weighted by
$e^{-\vd_\vM^T\widetilde{C}_{kl}\vd_\vM}$.  Here $\vd_\vM = \vs_k -
\vs_l - \vT_{\vM}$ is the shift difference between the two basis centers,
offset by the composite lattice translation vector $\vT_{\vM} =
(m_{1,x}L_x,\dots,m_{n,z}L_z)$, which displaces all $n$ electrons
simultaneously by integer multiples of the cell dimensions.  Because
$\widetilde{C}_{kl} = C_{kl}\otimes I_3$ is positive definite (inherited
from the positive-definiteness of the ECG matrices $A_k$ and
$A_l$), the image weight $e^{-\vd_\vM^T\widetilde{C}_{kl}\vd_\vM}$
decays exponentially as $|\vT_{\vM}|$ grows: contributions from distant
image cells are exponentially suppressed, guaranteeing absolute convergence
of every lattice sum and reducing it in practice to a finite shell of
images satisfying
$\vd_{\vM}^T\widetilde{C}_{kl}\vd_{\vM}\lesssim\chi^2_\mathrm{cut}$.

The availability of these matrix elements opens the way for applying
correlated Gaussians — with their systematically improvable accuracy — to a range of periodic
systems that have so far been accessible only to plane-wave or
Slater-type methods: hydrogen crystals in one, two, and three dimensions,
simple metals such as solid Li and Na (one valence electron per primitive
cell), trivalent solid Al, two-dimensional semimetals such as graphene
(two carbon atoms per cell), and other small-cell solids where a
pseudopotential reduces the active electron count to a tractable number.

For clarity and accessibility, definitions and results are presented
in the main text, while derivations are deferred to the appendices.

The paper is organized as follows.
In Sec.~\ref{sec:basis} we define the physical system, introduce the
shifted correlated Gaussian basis functions and their periodization
(Eq.~\eqref{eq:periodic_basis}), state the unfolding theorem and its
Bloch-phase generalization (Theorems~1 and~2), and collect the
composite quantities on which all matrix elements depend.
In Sec.~\ref{sec:results} we present closed-form expressions for all
matrix elements required by the variational problem: the overlap, the
kinetic energy, the Coulomb potential energy evaluated via three
independent routes (Ewald decomposition, direct neutral-shell sum, and
Dirac delta convolution), and the antisymmetrized total Hamiltonian
matrix element.
In Sec.~\ref{sec:h2chain} we specialize the general framework to a
one-dimensional hydrogen chain with two atoms per primitive cell,
working out the neutral-cell Coulomb matrix elements and Bloch twist
explicitly as a concrete illustration.
Sec.~\ref{sec:summary} summarizes the results in a unified table and
discusses computational efficiency, the charge-neutral simplification,
and prospective applications to hydrogen crystals, simple metals, and
two-dimensional materials.
The derivations are collected in the appendices:
Appendix~\ref{app:unfolding} proves the unfolding theorem and its
Bloch generalization;
Appendix~\ref{app:overlap} derives the overlap matrix element;
Appendix~\ref{app:kinetic} derives the kinetic energy matrix element;
Appendix~\ref{app:recip_coulomb} derives the reciprocal-space Coulomb
matrix elements via Fourier-modulated Gaussian integrals;
Appendix~\ref{app:real_coulomb} derives the real-space Coulomb matrix
elements via the $\mathrm{erfc}/r$ integral representation;
Appendix~\ref{app:neutral_coulomb} derives the neutral-cell screened
Coulomb result in closed form without an Ewald splitting parameter;
Appendix~\ref{app:delta} derives the Dirac delta matrix elements and
establishes their equivalence to the Coulomb matrix elements via
convolution;
Appendix~\ref{app:gaussians} collects the Gaussian integral identities
used throughout;
Appendix~\ref{app:gradients} gives gradient formulas for variational
optimization of the basis parameters;
Appendix~\ref{app:convergence} describes convergence acceleration for
diffuse basis functions via the Jacobi imaginary transformation and
the Poisson summation formula;
and Appendix~\ref{app:connection} establishes the exact correspondence
between the shifted correlated Gaussian parameterization used here and
the pair-correlation--single-particle Gaussian form common in the
literature.

\section{Physical System and Basis}
\label{sec:basis}

\subsection{Setup}

We consider a periodic solid containing $N_\mathrm{nuc}$ nuclei of
atomic numbers $\{Z_I\}_{I=1}^{N_\mathrm{nuc}}$ at fixed positions
$\{\vRR_I\}_{I=1}^{N_\mathrm{nuc}}$ inside an orthorhombic simulation cell
with lattice vectors $\vL=(L_x,L_y,L_z)$ and volume $\Omega=L_xL_yL_z$.
The system contains $n$ electrons with configuration
\begin{equation}
  \vrr = (\vrr_1,\vrr_2,\dots,\vrr_n)^T \in \mathbb{R}^{3n}.
\end{equation}
We work in atomic units ($\hbar=m_e=e=1$) throughout.

\subsection{Periodic Hamiltonian}
\label{sec:hamiltonian}

The electronic Hamiltonian under periodic boundary conditions has the form
\begin{equation}\label{eq:H}
  \hat{H} = \hat{T} + \hat{V}^{(1)} + \hat{V}^{(2)} + E_{NN},
\end{equation}
where the four terms are the kinetic energy, the one-body (electron–nuclear)
potential, the two-body (electron–electron) potential, and the constant
nuclear–nuclear repulsion energy, respectively.

\paragraph{Kinetic energy.}
\begin{equation}\label{eq:T_op}
  \hat{T} = -\frac{1}{2}\sum_{i=1}^n\nabla_{\vrr_i}^2.
\end{equation}
For the general mass-weighted form $\hat{T}=-\frac{1}{2}\sum_i m_i^{-1}\nabla_i^2$,
introduce the inverse-mass matrix $\Lambda=\mathrm{diag}(m_1^{-1},\dots,m_n^{-1})$;
the electronic case has $\Lambda=I_n$ in atomic units.

\paragraph{One-body potential.}
The one-body operator is a sum over single-electron terms,
\begin{equation}\label{eq:V1_op}
  \hat{V}^{(1)} = \sum_{i=1}^n v(\vrr_i),
\end{equation}
where $v(\vrr_i)$ is a local, lattice-periodic potential.  For the
electron–nuclear Coulomb attraction under PBC, each nucleus at $\vRR_I$
inside the reference cell is accompanied by periodic images at
$\vRR_I + \vn\cdot\vL$ for all $\vn\in\mathbb{Z}^3$, giving
\begin{equation}\label{eq:v_eN}
  v(\vrr_i) = -\sum_{I=1}^{N_\mathrm{nuc}} Z_I
  \sum_{\vn\in\mathbb{Z}^3}
  \frac{1}{|\vrr_i - \vRR_I - \vn\cdot\vL|},
\end{equation}
where $\vn\cdot\vL=(n_xL_x,n_yL_y,n_zL_z)$ translates the image by $n_x,n_y,n_z$
cells along each lattice direction.  This sum is conditionally convergent
and is evaluated via the Ewald decomposition (see Eq. \eqref{eq:ewald}), written
compactly as $|\vrr_i-\vRR_I|_\mathrm{Ewald}^{-1}$.  Other one-body terms,
such as an external electric field or a pseudopotential, enter through the
same structure.

\paragraph{Two-body potential.}
The two-body operator acts on pairs of electrons,
\begin{equation}\label{eq:V2_op}
  \hat{V}^{(2)} = \sum_{i=1}^{n-1}\sum_{j=i+1}^{n} w(\vrr_i,\vrr_j),
\end{equation}
where $w(\vrr_i,\vrr_j)$ is a symmetric, lattice-periodic pair potential.
For the electron–electron Coulomb repulsion, each pair displacement
$\vrr_i-\vrr_j$ interacts with all periodic images of the charge
density, so
\begin{equation}\label{eq:w_ee}
  w(\vrr_i,\vrr_j) =
  \sum_{\vn\in\mathbb{Z}^3}
  \frac{1}{|\vrr_i - \vrr_j - \vn\cdot\vL|},
\end{equation}
where the $\vn=\mathbf{0}$ term is the direct interaction within the
reference cell and $\vn\neq\mathbf{0}$ terms are the image contributions.
This sum is again conditionally convergent and regularized by Ewald
summation (see Eq. \eqref{eq:ewald}), written as $|\vrr_i-\vrr_j|_\mathrm{Ewald}^{-1}$.

\paragraph{Nuclear–nuclear energy.}
The nuclear positions $\{\vRR_I\}$ are fixed (Born–Oppenheimer
approximation), so the nuclear–nuclear Coulomb repulsion is the
classical electrostatic energy of the nuclear lattice summed over all
image cells,
\begin{equation}\label{eq:E_NN}
  E_{NN} = \frac{1}{2}
  \sum_{I,J=1}^{N_\mathrm{nuc}}
  \sum_{\vn\in\mathbb{Z}^3}{}^{\prime}
  \frac{Z_IZ_J}{|\vRR_I - \vRR_J - \vn\cdot\vL|},
\end{equation}
where the prime on the sum excludes the self-interaction term $I=J$,
$\vn=\mathbf{0}$.  This is the standard Madelung sum, which is
conditionally convergent and evaluated via Ewald summation
(see Eq. \eqref{eq:ewald}) as $|\vRR_I-\vRR_J|_\mathrm{Ewald}^{-1}$; it is a
constant that shifts all energy eigenvalues uniformly.

\subsection{Shifted correlated Gaussian basis functions}

A single (non-periodized) basis function is
\begin{equation}\label{eq:scg}
  \phi_k(\vrr) = \exp\!\Big[-(\vrr-\vs_k)^T\At_k\,(\vrr-\vs_k)\Big],
\end{equation}
where $A_k\in\mathbb{R}^{n\times n}$ is a symmetric positive-definite
\emph{correlation matrix} parameterized via its Cholesky factor,
$A_k=L_kL_k^T$; $\At_k=A_k\otimes I_3$ is the $3n\times 3n$
Kronecker-expanded matrix; and $\vs_k\in\mathbb{R}^{3n}$ is the shift
(centering) vector.  Off-diagonal elements of $A_k$ encode
electron–electron correlations explicitly.

\subsection{Periodized basis functions}

Because each electron must be periodic in the simulation cell, we introduce
the composite lattice translation
\begin{widetext}
\begin{equation}\label{eq:TM}
  \vT_{\vM}=\bigl(m_{1,x}L_x,m_{1,y}L_y,m_{1,z}L_z,\,\dots,\,
  m_{n,x}L_x,m_{n,y}L_y,m_{n,z}L_z\bigr),
\end{equation}
\end{widetext}
with $\vm_i\in\mathbb{Z}^3$, and the \emph{periodized} basis function
\begin{equation}\label{eq:periodic_basis}
  \Phi_k(\vrr) = \sum_{\vM\in\mathbb{Z}^{3n}}\phi_k(\vrr-\vT_{\vM}).
\end{equation}
For every composite lattice vector $T_N$ ($N\in\mathbb{Z}^{3n}$),
\begin{equation}\label{eq:periodic}
\Phi_k(\bm{r}+T_N) = \Phi_k(\bm{r}).
\end{equation}

To compute the electronic band structure $E_n(\bm{k}_B)$ one must
evaluate matrix elements at many Bloch wave vectors
$\bm{k}_B$ in the first Brillouin zone.  The
plain periodized function~\eqref{eq:periodic_basis} corresponds to
$\bm{k}_B=\bm{0}$ (the $\Gamma$-point).  The generalization
is the Bloch-twisted basis function (it will be discussed in Theorem~2
of this paper):
\begin{equation}\label{eq:Phi_kB}
\Phi_k^{(\bm{k}_B)}(\bm{r})
= \sum_{M\in\mathbb{Z}^{3n}}
e^{i\bm{k}_B\cdot T_M}\,\varphi_k(\bm{r}-T_M).
\end{equation}
Each image is now weighted by the Bloch phase
$e^{i\bm{k}_B\cdot T_M}$, a unit complex number. 
The function $\Phi_k^{(\bm{k}_B)}$ satisfies the Bloch condition
under every composite lattice translation:
\begin{equation}\label{eq:bloch}
\Phi_k^{(\bm{k}_B)}(\bm{r}+T_N)
= e^{i\bm{k}_B\cdot T_N}\,\Phi_k^{(\bm{k}_B)}(\bm{r})
\quad \forall\, N\in\mathbb{Z}^{3n}.
\end{equation}

\subsection{General matrix element and variational problem}
\label{sec:matelem}

The variational method approximates eigenstates of $\hat{H}$ as linear
combinations of the periodized basis functions,
\begin{equation}\label{eq:wavefunction}
  \Psi(\vrr) \approx \sum_{k=1}^K c_k\,\Phi_k(\vrr),
\end{equation}
and determines the coefficients $\{c_k\}$ by the Ritz principle.  This
requires evaluating, for every operator $\hat{O}$ appearing in $\hat{H}$
and for every pair of basis indices $(k,l)$, the \emph{periodic matrix element}
\begin{equation}\label{eq:matelem}
O_{kl} \equiv \langle\Phi_k|\hat{O}|\Phi_l\rangle
  = \int_{\Omega^n}\Phi_k^*(\vrr)\,\hat{O}\,\Phi_l(\vrr)\,d\vrr.
\end{equation}
The integration domain is the n-electron simulation cell $\Omega^n$, and
$\hat{O}$ ranges over the identity (overlap), $\hat{T}$, $\hat{V}^{(1)}$,
and $\hat{V}^{(2)}$.  

By the unfolding theorem (see Theorem~1 below), the full periodic
matrix element at the $\Gamma$-point is
\begin{equation}\label{eq:Okl0}
O_{kl} = \langle\Phi_k|\hat{O}|\Phi_l\rangle
= \sum_{M\in\mathbb{Z}^{3n}} O_{kl}(M),
\end{equation}
where 
\begin{equation}\label{eq:Oklm}
O_{kl}(M)
= \int_{\mathbb{R}^{3n}}
\varphi_k(\bm{r})\,\hat{O}\,\varphi_l(\bm{r}-T_M)\,d\bm{r},
\end{equation}
and now the integration is over all of $\mathbb{R}^{3n}$. 
Note that the cell integral over $\Phi_k$,$\Phi_l$ 
and the all-space integral over $\varphi_k, \varphi_l$
are two different but exactly equivalent ways of writing the
same number. The first is the natural physical definition; the second
is what makes the Gaussian integrals analytically tractable, since
individual Gaussians $\varphi_k$ 
are not periodic and their integrals over a finite cell would not
have closed forms. The unfolding theorem is precisely the bridge that
makes the analytic evaluation possible.

Assembling all pairs into the $K\times K$
Hamiltonian matrix $\mathbf{H}=(H_{kl})$ and overlap matrix
$\mathbf{S}=(S_{kl})$, the variational coefficients satisfy the
\emph{generalized eigenvalue problem}
\begin{equation}\label{eq:gev}
  \mathbf{H}\,\mathbf{c} = E\,\mathbf{S}\,\mathbf{c},
\end{equation}
whose lowest eigenvalue $E_0$ is an upper bound on the exact ground-state
energy.  

\paragraph{Bloch matrix elements.}
Because $\Phi_k^{(\bm{k}_B)}$ satisfies~\eqref{eq:bloch}, the
Hamiltonian and overlap matrices
\[
H_{kl}(\bm{k}_B) = \langle\Phi_k^{(\bm{k}_B)}|\hat
H|\Phi_l^{(\bm{k}_B)}\rangle,
\quad
S_{kl}(\bm{k}_B) =
\langle\Phi_k^{(\bm{k}_B)}|\Phi_l^{(\bm{k}_B)}\rangle
\]
are \emph{complex Hermitian} for $\bm{k}_B\neq\bm{0}$.  By the
unfolding theorem (Theorem~1 of the paper) they reduce to
\begin{equation}
O_{kl}(\bm{k}_B)
= \sum_{M\in\mathbb{Z}^{3n}} e^{i\bm{k}_B\cdot T_M}
O_{kl}(M)
\end{equation}
i.e.\ the same real-space integrals as at the
$\Gamma$-point, but weighted by Bloch phases. 
This means no new integrals need to be
computed for any $\mathbf{k}_B$
point — the Bloch band structure is just a phase-weighted Fourier
sum over the same image integrals computed once at the $\Gamma$-point.

\paragraph{Band structure.}
Solving the generalized eigenvalue problem
\[
H(\bm{k}_B)\,\bm{c} = E(\bm{k}_B)\,S(\bm{k}_B)\,\bm{c}
\]
at many $\bm{k}_B$-points across the first
Brillouin zone yields the
electronic band structure $E_n(\bm{k}_B)$.  At
$\bm{k}_B=\bm{0}$ all
Bloch phases are unity and the real eigenvalue
problem of Eq.\,\ref{eq:gev} is recovered.

\subsection{Composite quantities for a bra-ket pair $(k,l)$}
\label{sec:composite}

All matrix elements depend on the following quantities.  The combined
nonlinear parameter matrix and its Kronecker form are
\begin{equation}\label{eq:Akl}
  A_{kl}=A_k+A_l, \qquad \At_{kl}=A_{kl}\otimes I_3.
\end{equation}
The \emph{reduced} (harmonic-mean) matrix is
\begin{equation}\label{eq:Ckl}
  C_{kl}=A_k\,A_{kl}^{-1}\,A_l, \qquad \widetilde{C}_{kl}=C_{kl}\otimes I_3.
\end{equation}
For image index $\vM$, the \emph{shift difference} and the \emph{combined
Gaussian center} are
\begin{align}
  \vd_{\vM} &= \vs_k-\vs_l-\vT_{\vM}, \label{eq:dM}\\
  \bar{\vrr}_{\vM} &= \At_{kl}^{-1}
    \bigl(\At_k\,\vs_k+\At_l(\vs_l+\vT_{\vM})\bigr). \label{eq:rbar}
\end{align}
The common prefactor arising from a $3n$-dimensional Gaussian integral is
\begin{equation}\label{eq:calS}
  \mathcal{S}_{kl} = \frac{\pi^{3n/2}}{(\det A_{kl})^{3/2}},
\end{equation}
where we used $\det(A_{kl}\otimes I_3)=(\det A_{kl})^3$.

We also define the image weight
\begin{equation}\label{eq:omegaM}
  \omega_{\vM} = \exp\!\bigl[-\vd_{\vM}^T\,\widetilde{C}_{kl}\,\vd_{\vM}\bigr],
\end{equation}
and the kinetic-energy auxiliary matrix
\begin{equation}\label{eq:Bkl}
  B_{kl}^{(\Lambda)} = C_{kl}\,\Lambda\,C_{kl}
  = A_k A_{kl}^{-1}A_l\,\Lambda\,A_l A_{kl}^{-1}A_k,
\end{equation}
where $\Lambda=\mathrm{diag}(m_1^{-1},\dots,m_n^{-1})$ is the inverse-mass
matrix.

For the Coulomb terms we need the \emph{effective pair widths}.  For an
electron-electron pair $(i,j)$,
\begin{equation}\label{eq:sigma_ee}
  \sigma_{ij,s}^2 = (A_{kl}^{-1})_{ii}+(A_{kl}^{-1})_{jj}
                   -2(A_{kl}^{-1})_{ij},
\end{equation}
and for a single electron $i$,
\begin{equation}\label{eq:sigma_en}
  \sigma_{i}^2 = (A_{kl}^{-1})_{ii}.
\end{equation}
The coordinate projectors from $3n$-space to the physical 3D pair
displacement are
\begin{equation}\label{eq:projectors}
  \vP_{ij} = (\ve_i-\ve_j)\otimes I_3, \qquad \vP_i = \ve_i\otimes I_3,
\end{equation}
where $\ve_i$ is the $i$-th standard basis vector in $\mathbb{R}^n$.

\subsection{Unfolding theorem for periodic operators}
\label{sec:unfolding}

Inserting two periodized basis functions into the matrix element of an
operator $\hat{O}$ always produces a double sum over image indices.  The
following theorem, whose proof is given in Appendix~\ref{app:unfolding},
reduces this double sum to a single one under a natural periodicity
condition on $\hat{O}$.

We say that $\hat{O}$ is \emph{lattice-periodic} if its integral kernel
$K_{\hat{O}}(\vrr,\vrr')=\langle\vrr|\hat{O}|\vrr'\rangle$ satisfies
\begin{equation}\label{eq:kernel_periodic}
  K_{\hat{O}}(\vrr+\vT_{\vM},\vrr'+\vT_{\vM})
  = K_{\hat{O}}(\vrr,\vrr'),
  \quad \forall\,\vM\in\mathbb{Z}^{3n}.
\end{equation}
All operators of physical interest are of this type: local potentials
$V(\vrr)$ with $V(\vrr+\vT_{\vM})=V(\vrr)$, differential operators such as
$\nabla^2$, and non-local operators such as the exchange operator.

\textbf{Theorem 1 (Unfolding).}
\textit{Let $\hat{O}$ be lattice-periodic in the sense of
Eq.~\eqref{eq:kernel_periodic}.  Then}
\begin{align}
  O_{kl} &\equiv \int_{\Omega^n}\Phi_k^*(\vrr)\,\hat{O}\,\Phi_l(\vrr)\,d\vrr
  \notag\\
  &= \sum_{\vM\in\mathbb{Z}^{3n}}
     \int_{\mathbb{R}^{3n}}\phi_k(\vrr)\,\hat{O}\,\phi_l(\vrr-\vT_{\vM})\,d\vrr.
  \label{eq:unfolding}
\end{align}

\noindent
The theorem states that for any lattice-periodic operator the double image
sum collapses: one index is absorbed by promoting the cell integral to an
all-space integral, and the remaining index $\vM=\vM'-\vM_\mathrm{bra}$
measures the \emph{relative} image offset between ket and bra.

\textbf{Corollary 1a (Local operators).}  \textit{If $\hat{O}=V(\vrr)$ is
a multiplicative, cell-periodic potential, Eq.~\eqref{eq:unfolding} reduces
to}
\begin{equation}\label{eq:unfolding_local}
  O_{kl} = \sum_{\vM}\int_{\mathbb{R}^{3n}}
  \phi_k(\vrr)\,V(\vrr)\,\phi_l(\vrr-\vT_{\vM})\,d\vrr.
\end{equation}

\textbf{Corollary 1b (Differential operators).}  \textit{If
$\hat{O}=\hat{D}(\nabla)$ is any polynomial in $\nabla_\vrr$ (e.g.\
$-\nabla^2/2$), which trivially commutes with translations, then
Eq.~\eqref{eq:unfolding} holds and the action of $\hat{D}$ is taken on
$\phi_l(\vrr-\vT_{\vM})$ inside the $\mathbb{R}^{3n}$ integral.}

\textbf{Corollary 1c (Non-local, cell-periodic operators).}  \textit{If
$\hat{O}$ is non-local with kernel $K(\vrr,\vrr')$ satisfying
Eq.~\eqref{eq:kernel_periodic}, then}
\begin{align}\label{eq:unfolding_nonlocal}
  O_{kl} = \sum_{\vM}\int_{\mathbb{R}^{3n}}\!\int_{\mathbb{R}^{3n}}
  \phi_k(\vrr)\,K(\vrr,\vrr')\,\phi_l(\vrr'-\vT_{\vM})\,d\vrr\,d\vrr'.
\end{align}

\textbf{Corollary 1d (Ewald-summed operators).}  \textit{If $\hat{O}$
acts through the periodized Coulomb kernel
$\sum_{\vn\in\mathbb{Z}^3}\nu(\vrr,\vn)$
(where $\nu(\vrr,\vn)$ depends on a real-space shell $\vn$ and on
inter-particle distances), then the unfolding produces a triple sum:}
\begin{align}\label{eq:unfolding_ewald}
  O_{kl} = \sum_{\vn\in\mathbb{Z}^3}\sum_{\vM\in\mathbb{Z}^{3n}}
  \int_{\mathbb{R}^{3n}}
  \phi_k(\vrr)\,\nu(\vrr,\vn)\,\phi_l(\vrr-\vT_{\vM})\,d\vrr,
\end{align}
\textit{with the reciprocal-space (Fourier) part of the Ewald sum yielding
only the $\vM$-sum.}

\textbf{Bloch generalization (Theorem 2).}
\textit{At a Bloch wave vector $\mathbf{k}_\mathrm{B}$, the periodized
basis functions carry a phase,}
\begin{equation}
  \Phi_k^{(\mathbf{k}_\mathrm{B})}(\vrr)
  =\sum_{\vM}e^{i\mathbf{k}_\mathrm{B}\cdot\vT_{\vM}}\phi_k(\vrr-\vT_{\vM}).
\end{equation}
\textit{If $\hat{O}$ satisfies Eq.~\eqref{eq:kernel_periodic}, then}
\begin{align}\label{eq:unfolding_bloch}
  O_{kl}^{(\mathbf{k}_\mathrm{B})}
  &\equiv\int_{\Omega^n}
    \bigl[\Phi_k^{(\mathbf{k}_\mathrm{B})}(\vrr)\bigr]^*
    \hat{O}\,\Phi_l^{(\mathbf{k}_\mathrm{B})}(\vrr)\,d\vrr
  \notag\\
  &= \sum_{\vM}e^{i\mathbf{k}_\mathrm{B}\cdot\vT_{\vM}}
     \int_{\mathbb{R}^{3n}}
     \phi_k(\vrr)\,\hat{O}\,\phi_l(\vrr-\vT_{\vM})\,d\vrr.
\end{align}
\textit{The Bloch factor $e^{i\mathbf{k}_\mathrm{B}\cdot\vT_{\vM}}$ weights
each image by a phase; the $\Gamma$-point result $\mathbf{k}_\mathrm{B}=\mathbf{0}$
recovers Theorem~1.}

The derivation is given in Appendix~\ref{app:unfolding}.

\subsection{Periodic Coulomb (Ewald) potential}
\label{sec:ewald}

Under periodic boundary conditions all Coulomb sums are conditionally
convergent and are regularized by Ewald summation.  For two charges
separated by $\vu$ inside the simulation cell,
\begin{align}
  \frac{1}{|\vu|_\mathrm{Ewald}} =&
  \sum_{\vn\in\mathbb{Z}^3}
    \frac{\mathrm{erfc}(\kappa|\vu+\vn\cdot\vL|)}{|\vu+\vn\cdot\vL|}
  \notag\\
  &+\frac{4\pi}{\Omega}\sum_{\vG\neq\mathbf{0}}
    \frac{e^{-G^2/4\kappa^2}}{G^2}\,e^{i\vG\cdot\vu}
  -\frac{\pi}{\kappa^2\Omega},\label{eq:ewald}
\end{align}
where $\kappa$ is the Ewald splitting parameter, $\vn\in\mathbb{Z}^3$
labels real-space lattice images with translation $\vn\cdot\vL =
(n_xL_x, n_yL_y, n_zL_z)$, and $\vG = 2\pi(n_x/L_x, n_y/L_y,
n_z/L_z)$ are the reciprocal lattice vectors with $G=|\vG|$ and cell
volume $\Omega = L_xL_yL_z$.

The key idea of the Ewald split is to decompose the slowly decaying $1/r$
interaction into two rapidly convergent parts via the identity
$1 = \mathrm{erfc}(\kappa r) + \mathrm{erf}(\kappa r)$.  The
complementary error function $\mathrm{erfc}(\kappa|\vu+\vn\cdot\vL|)$
decays as $e^{-\kappa^2|\vn\cdot\vL|^2}$ for large $|\vn|$, making the
real-space sum absolutely convergent with a shell radius proportional to
$1/\kappa$.  The error-function remainder $\mathrm{erf}(\kappa r)/r$ is
smooth and periodic, so its Fourier series converges rapidly: the
reciprocal-space coefficients carry a factor $e^{-G^2/4\kappa^2}$, which
damps high-$G$ shells at a rate set by $1/\kappa$.  The last term
$-\pi/(\kappa^2\Omega)$ is the $\vG=\mathbf{0}$ (uniform background)
correction required for a charged system.  Increasing $\kappa$ accelerates
the real-space convergence at the cost of slower reciprocal-space
convergence, and vice versa; the optimal $\kappa\sim\pi/L$ balances the
two.

\section{Matrix Elements: Results}
\label{sec:results}

We evaluate $O_{kl}=\langle\Phi_k|\hat{O}|\Phi_l\rangle$
(Eq.~\eqref{eq:matelem}) for each operator in $\hat{H}$, using the
notation of Sec.~\ref{sec:composite}.
Derivations are collected in the appendices.

\subsection{Overview: operators and their matrix elements}
\label{sec:overview}

The variational calculation of the band structure of a periodic solid
requires matrix elements of four classes of operators.

The \emph{overlap} $S_{kl}=\langle\Phi_k|\Phi_l\rangle$ enters the
generalized eigenvalue problem \eqref{eq:gev} and measures the
non-orthogonality of the periodized Gaussian basis.

The \emph{kinetic energy} $T_{kl}=\langle\Phi_k|\hat{T}|\Phi_l\rangle$
is the dominant one-body contribution at short range and determines the
curvature of the dispersion bands.

The \emph{Coulomb potential energy} — electron–electron, electron–nuclear,
and nuclear–nuclear — is the physically central and technically most
demanding part of the calculation.  It is computed in three independent
ways, all of which yield identical results and serve as mutual consistency
checks.

\begin{enumerate}
\item \textit{Ewald decomposition} (Secs.~\ref{sec:ewald}--\ref{sec:results}).
The periodic $1/r$ potential is split into a short-range complementary-error-function
part evaluated in real space and a long-range smooth part evaluated in
reciprocal space, following Eq.~\eqref{eq:ewald}.  This approach is
valid for any charge configuration (neutral or charged cell) and is the
standard method for periodic electronic-structure calculations.

\item \textit{Direct neutral-cell sum} (Sec.~\ref{sec:neutral_coulomb}).
When the simulation cell is charge-neutral the lattice sum over $1/r$
converges absolutely when shells are grouped by charge neutrality.  Each
matrix element reduces analytically to a screened Coulomb potential
$\mathrm{erf}(R/\sigma)/R$, with no Ewald splitting parameter $\kappa$
required.  This formulation is algebraically simpler and avoids any
tuning of $\kappa$.

\item \textit{Dirac delta convolution} (Sec.~\ref{sec:delta}).
The $1/r$ kernel is the convolution of a Dirac delta density with the
Green's function of the Laplacian.  The matrix element of
$\delta^{(3)}(\vrr_i-\vrr_j)$ gives the pair-contact density between
electrons $i$ and $j$, and the $1/r$ Coulomb matrix element is recovered
by integrating this density against $1/|\mathbf{u}|$:
\begin{align}
  V^{(ee)}_{kl} &= \sum_{i<j}
  \int_{\mathbb{R}^3}
  \frac{\langle\Phi_k|\delta^{(3)}(\vrr_i-\vrr_j-\mathbf{u})|\Phi_l\rangle}
       {|\mathbf{u}|}\,d^3\mathbf{u}, \label{eq:Vee_from_delta}
\end{align}
and analogously for the electron–nuclear term with
$\delta^{(3)}(\vrr_i-\mathbf{S})$ integrated against
$1/|\mathbf{S}-\vRR_I|$.  Evaluating the convolution with the identity
$\int e^{-|\mathbf{t}-\mathbf{R}|^2/\sigma^2}/|\mathbf{t}|\,d^3\mathbf{t}
=\pi^{3/2}\sigma^{-1}\mathrm{erf}(|\mathbf{R}|/\sigma)/|\mathbf{R}|$
reproduces the neutral-cell result exactly, confirming the equivalence of
all three approaches.
\end{enumerate}

Beyond the Coulomb matrix elements, the Dirac delta matrix elements have
independent physical significance.  The pair-contact density
$\langle\Phi_k|\delta^{(3)}(\vrr_i-\vrr_j)|\Phi_l\rangle$ enters the
electron–electron cusp condition and the pair-correlation function
$g(r)$.  The single-electron density operator
$\delta^{(3)}(\vrr_i-\mathbf{S})$ evaluated at $\mathbf{S}=\vRR_I$ gives
the contact density at nucleus $I$, which determines the Fermi contact
hyperfine coupling.  More generally, the same convolution identity
\eqref{eq:Vee_from_delta} enables matrix elements of \emph{any} potential
that can be expressed as a convolution with the delta density:
\begin{equation}\label{eq:general_potential}
  \langle\Phi_k|V(\vrr_i-\vrr_j)|\Phi_l\rangle
  = \int_{\mathbb{R}^3}
    \langle\Phi_k|\delta^{(3)}(\vrr_i-\vrr_j-\mathbf{u})|\Phi_l\rangle\,
    V(\mathbf{u})\,d^3\mathbf{u},
\end{equation}
so that Yukawa, Gaussian, and other model pair potentials can be handled
within the same framework simply by replacing the $1/|\mathbf{u}|$ kernel
with the appropriate $V(\mathbf{u})$.  The delta matrix elements thus act
as a universal building block for potential energy evaluation.

\subsection{Overlap}
The overlap of the basis functions is derived in Appendix
\ref{app:overlap}:
\begin{equation}\label{eq:Skl}
  \boxed{S_{kl} = \mathcal{S}_{kl}\sum_{\vM\in\mathbb{Z}^{3n}}\omega_{\vM}.}
\end{equation}
The sum converges exponentially because $C_{kl}$ is positive definite.  In
practice only images satisfying
$\vd_{\vM}^T\widetilde{C}_{kl}\vd_{\vM}\lesssim\chi^2_\mathrm{cut}$ (with
$\chi^2_\mathrm{cut}\approx 20$–$30$ for double precision) contribute.

\subsection{Kinetic energy}

Let $\Lambda=\mathrm{diag}(m_1^{-1},\dots,m_n^{-1})$ and define
$B_{kl}^{(\Lambda)}$ as in Eq.~\eqref{eq:Bkl}.  Then
\begin{equation}\label{eq:Tkl}
  \boxed{T_{kl} = \frac{1}{2}\,\mathcal{S}_{kl}
  \sum_{\vM}\omega_{\vM}
  \Bigl[6\,\Tr(\Lambda C_{kl})
       -4\,\vd_{\vM}^T(B_{kl}^{(\Lambda)}\otimes I_3)\vd_{\vM}\Bigr].}
\end{equation}
For equal electron masses $\Lambda=\lambda I_n$, one has
$B_{kl}^{(\Lambda)}=\lambda C_{kl}^2$ and $\Tr(\Lambda C_{kl})=\lambda\Tr(C_{kl})$.
The details of the derivations are in Appendix \ref{app:kinetic}.

\subsection{Coulomb potential energy}

The full Coulomb Hamiltonian is $V=V_{eN}+V_{ee}+V_{NN}$.  Each term is
evaluated via the Ewald decomposition \eqref{eq:ewald}.

\subsubsection{Electron--electron reciprocal-space term}
The electron--electron reciprocal-space matrix element is
(see Appendix \ref{app:recip_coulomb}
\begin{multline}\label{eq:Vee_G}
  V_{kl}^{(ee,\vG)} =
  \frac{4\pi}{\Omega}\,\mathcal{S}_{kl}
  \sum_{i=1}^{n-1}\sum_{j=i+1}^{n}\sum_{\vG\neq\mathbf{0}}
  \frac{e^{-G^2/4\kappa^2}}{G^2}\\
  \times\sum_{\vM}\omega_{\vM}\,
  e^{\,i\vG\cdot\vP_{ij}^T\bar{\vrr}_{\vM}-\sigma_{ij,s}^2 G^2/4}.
\end{multline}
The sum over reciprocal lattice vectors $\vG$ converges rapidly: the
factor $e^{-G^2/4\kappa^2}$ from the Ewald decomposition damps
contributions from large-$|\vG|$ shells at a Gaussian rate, while the
additional factor $e^{-\sigma_{ij,s}^2G^2/4}$ arising from the Gaussian
basis provides further exponential suppression controlled by the effective
pair width $\sigma_{ij,s}$.  In practice only a sphere of reciprocal
shells with $G^2 \lesssim 4\kappa^2\ln(1/\varepsilon)$ (where
$\varepsilon$ is the desired precision) is required.  The image sum
$\sum_{\vM}\omega_{\vM}(\cdots)$ converges for the same reason as the
overlap: $\omega_{\vM}$ decays exponentially with $|\vT_{\vM}|^2$.

\subsubsection{Electron--electron real-space term}

For the real-space sum we define the \emph{$t$-augmented} nonlinear
parameter matrix
and its determinant ratio,
\begin{align}
  A_{kl}^{(t,ij)} &= A_{kl}+t^2(\ve_i-\ve_j)(\ve_i-\ve_j)^T,
  \label{eq:Atij}\\
  \det A_{kl}^{(t,ij)} &= \det A_{kl}\cdot(1+t^2\sigma_{ij,s}^2),
  \label{eq:detlem}
\end{align}
where the second identity follows from the matrix determinant lemma.  The
reduced quadratic form for image $\vM$, real-space shell $\vn$, and
auxiliary variable $t$ is
\begin{align}
  \mathcal{Q}_{\vM}^{(t,ij,\vn)} &=
  \vd_{\vM}^T\widetilde{C}_{kl}^{(t,ij)}\vd_{\vM}
  +\frac{t^2|\vn\cdot\vL+\bar{u}_{ij,\vM}|^2}{1+t^2\sigma_{ij,s}^2},
  \label{eq:Qee}
\end{align}
where $\bar{u}_{ij,\vM}=\bar{\vrr}_{\vM,i}-\bar{\vrr}_{\vM,j}$ is the
mean pair displacement (the difference of the $i$-th and $j$-th 3D blocks
of $\bar{\vrr}_{\vM}$), and $\widetilde{C}_{kl}^{(t,ij)}$ is computed
with $A_{kl}^{(t,ij)}$ in place of $A_{kl}$.  Then (see Appendix
\ref{app:real_coulomb}) 
\begin{multline}\label{eq:Vee_real}
  V_{kl}^{(ee,\mathrm{real})} =
  \frac{2}{\sqrt{\pi}}\,\mathcal{S}_{kl}
  \sum_{i=1}^{n-1}\sum_{j=i+1}^{n}\sum_{\vn\in\mathbb{Z}^3}\sum_{\vM}
  \\
  \times\int_\kappa^\infty
    \frac{\exp\!\bigl[-\mathcal{Q}_{\vM}^{(t,ij,\vn)}\bigr]}
         {(1+t^2\sigma_{ij,s}^2)^{3/2}}\,dt.
\end{multline}
The $t$-integral is evaluated by Gauss–Legendre quadrature after mapping
$[\kappa,\infty)\to(0,1]$ via $t=\kappa/u$;
$N_\mathrm{quad}\sim 10$–$20$ points typically achieve double-precision
accuracy.  The real-space sum over lattice shells $\vn\in\mathbb{Z}^3$
converges because the Ewald $\mathrm{erfc}$ factor damps the integrand:
for large $|\vn\cdot\vL|$ the pair displacement $|\vn\cdot\vL +
\bar{u}_{ij,\vM}| \sim |\vn\cdot\vL|$, so the factor
$e^{-t^2|\vn\cdot\vL+\bar{u}_{ij,\vM}|^2}$ (which appears in
$e^{-\mathcal{Q}_\vM^{(t,ij,\vn)}}$ via the second term of
Eq.~\eqref{eq:Qee}) decays as $e^{-\kappa^2|\vn\cdot\vL|^2}$ at the
lower limit $t=\kappa$, giving absolute convergence with a shell radius
$\sim 1/\kappa$.  The image sum over $\vM$ converges exponentially
through $\omega_{\vM}$, as for all other matrix elements.

\subsubsection{Electron--nuclear terms}

The electron–nuclear interaction is obtained from the electron–electron
expressions by the replacements
\begin{equation}\label{eq:eN_replace}
  \ve_i-\ve_j\to\ve_i, \quad
  \sigma_{ij,s}^2\to\sigma_i^2, \quad
  \bar{u}_{ij,\vM}\to\bar{\vrr}_{\vM,i}-\vRR_I,
\end{equation}
together with the charge factor $-Z_I$ and a sum over nuclei.

Reciprocal space:
\begin{multline}\label{eq:VeN_G}
  V_{kl}^{(eN,\vG)} =
  -\frac{4\pi}{\Omega}\,\mathcal{S}_{kl}
  \sum_{i=1}^{n}\sum_{I=1}^{N_\mathrm{nuc}} Z_I
  \sum_{\vG\neq\mathbf{0}}
  \frac{e^{-G^2/4\kappa^2}}{G^2}\\
  \times\sum_{\vM}\omega_{\vM}\,
  e^{\,i\vG\cdot(\bar{\vrr}_{\vM,i}-\vRR_I)-\sigma_i^2 G^2/4}.
\end{multline}

Real space (with $\mathcal{Q}_{\vM}^{(t,i,I,\vn)}$ obtained from
Eq.~\eqref{eq:Qee} via the substitutions \eqref{eq:eN_replace}):
\begin{multline}\label{eq:VeN_real}
  V_{kl}^{(eN,\mathrm{real})} =
  -\frac{2}{\sqrt{\pi}}\,\mathcal{S}_{kl}
  \sum_{i=1}^{n}\sum_{I=1}^{N_\mathrm{nuc}}Z_I\sum_{\vn\in\mathbb{Z}^3}\sum_{\vM}
  \\
  \times\int_\kappa^\infty
    \frac{\exp\!\bigl[-\mathcal{Q}_{\vM}^{(t,i,I,\vn)}\bigr]}
         {(1+t^2\sigma_i^2)^{3/2}}\,dt.
\end{multline}

\subsubsection{Nuclear--nuclear interaction}

Because $V_{NN}$ is independent of electronic coordinates it is proportional
to the overlap:
\begin{equation}\label{eq:VNN}
  V_{kl}^{(NN)} = E_\mathrm{Madelung}\cdot S_{kl},
\end{equation}
where
$E_\mathrm{Madelung}=\frac{1}{2}{\sum_{I=1}^{N_\mathrm{nuc}}
\sum_{J=1}^{N_\mathrm{nuc}}}^{\prime}\,Z_IZ_J/|\vRR_I-\vRR_J|_\mathrm{Ewald}$
is the standard Ewald nuclear-repulsion energy (prime excludes $I=J$ in the
same cell):

\begin{align}
E_{\mathrm{Madelung}} 
&= \frac{1}{2} \sum_{I,J}^{\prime} Z_I Z_J 
\left[
\sum_{\mathbf{n} \in \mathbb{Z}^3}
\frac{\mathrm{erfc}\!\left(\kappa |R_I - R_J +
\mathbf{n}\cdot\mathbf{L}|\right)}
{|R_I - R_J + \mathbf{n}\cdot\mathbf{L}|}
\right.
\nonumber \\
&\quad
+ \frac{4\pi}{\Omega}
\sum_{\mathbf{G} \neq \mathbf{0}}
\frac{e^{-G^2/4\kappa^2}}{G^2}
e^{i\mathbf{G}\cdot(R_I - R_J)}
\nonumber \\
&\quad
\left.
- \frac{\pi}{\kappa^2 \Omega}
- \frac{2\kappa}{\sqrt{\pi}}\,\delta_{IJ}
\right]
\label{eq:madelung_full}
\end{align}

\subsubsection{Ewald self-energy correction}

The Ewald decomposition introduces an unphysical electronic self-interaction,
\begin{equation}\label{eq:Vself}
  V_{kl}^{(\mathrm{self})} =
  \Bigl(-\frac{\kappa}{\sqrt{\pi}}\,n
  -\frac{\pi n(n-1)}{2\kappa^2\Omega}\Bigr)S_{kl}.
\end{equation}

\subsubsection{Cancellation of divergences and the neutral-cell alternative}

The individual terms $V_{kl}^{(ee,\vG)}$, $V_{kl}^{(eN,\vG)}$, and the
$\vG=\mathbf{0}$ background correction in Eq.~\eqref{eq:ewald} each
diverge for a charged or uniform-background system taken in isolation.
These divergences cancel exactly in the total Coulomb matrix element: the
repulsive electron–electron and attractive electron–nuclear reciprocal-space
contributions combine with the self-energy correction
$V_{kl}^{(\mathrm{self})}$ and the Madelung term $V_{kl}^{(NN)}$ to yield
a finite, $\kappa$-independent result for the physical (charge-neutral)
combination $V_{ee}+V_{eN}+V_{NN}$.  The Ewald splitting parameter
$\kappa$ merely controls the partition of work between the real-space and
reciprocal-space sums; the total is invariant.

For a charge-neutral simulation cell an alternative, simpler route is
available: the bare $1/r$ lattice sum converges absolutely when grouped
into neutral shells, and the matrix elements reduce to the closed-form
screened Coulomb expressions of Sec.~\ref{sec:neutral_coulomb} without any
$\kappa$-dependent regularization.  In this case no self-energy correction
is needed and all individual terms are finite from the outset, providing a
useful independent check on the Ewald calculation.

\subsection{Neutral Coulomb potential without Ewald summation}
\label{sec:neutral_coulomb}

For a \emph{neutral} simulation cell — total electron plus nuclear charge equal
to zero — the periodic $1/r$ sum is absolutely convergent and may be evaluated
directly, without Ewald decomposition.  This is the relevant case for any system
where the charge-neutrality condition removes the conditional-convergence
problem.

The convergence mechanism is transparent: when charges are grouped into
neutral shells (reference cell plus successive image shells), each shell
contributes a dipole-like field that falls off faster than $1/r^2$, making
the shell sum absolutely convergent.  Concretely, the image-$\vM$
contribution to every matrix element is weighted by $\omega_{\vM} =
e^{-\vd_{\vM}^T\widetilde{C}_{kl}\vd_{\vM}}$, which already provides
exponential damping in $|\vT_{\vM}|^2$; for a neutral cell the
bare-Coulomb $\mathrm{erf}(R/\sigma)/R$ kernel is additionally bounded for
all $R\geq 0$ (the $R\to 0$ limit is $2/(\sigma\sqrt{\pi})$), so no
additional $\kappa$-regularization is needed.

After the unfolding theorem (Theorem~1) with the bare Coulomb kernel, (which is
lattice-periodic for a neutral cell), each image term involves the integral
\begin{equation}\label{eq:bare_coul_int}
  \mathcal{J}_{\vM}^{(ij)} = \int_{\mathbb{R}^{3n}}
  \frac{\phi_k(\vrr)\,\phi_l(\vrr-\vT_{\vM})}
       {|\vP_{ij}^T\vrr|}\,d\vrr.
\end{equation}
Combining the Gaussians and using the Boys-function identity derived in
Appendix~\ref{app:neutral_coulomb}, the result is:

\begin{equation}\label{eq:J_ij_M}
  \mathcal{J}_{\vM}^{(ij)} = \omega_{\vM}\,\mathcal{S}_{kl}\,
  \frac{\mathrm{erf}\!\bigl(|\bar{u}_{ij,\vM}|/\sigma_{ij,s}\bigr)}{|\bar{u}_{ij,\vM}|},
\end{equation}
where $\bar{u}_{ij,\vM}=\bar{\vrr}_{\vM,i}-\bar{\vrr}_{\vM,j}$ is the mean
electron pair displacement (Eq.~\eqref{eq:rbar}), $\sigma_{ij,s}$ is the
effective pair width (Eq.~\eqref{eq:sigma_ee}), and the limit
$\lim_{R\to 0}\mathrm{erf}(R/\sigma)/R = 2/(\sigma\sqrt{\pi})$ resolves
the apparent singularity at $\bar{u}_{ij,\vM}=\mathbf{0}$.

The physical interpretation is transparent: $\mathrm{erf}(R/\sigma)/R$ is a
screened Coulomb potential with range $\sigma_{ij,s}$ — the Gaussian
smearing of the pair coordinate replaces the bare $1/R$ singularity with a
smooth, finite kernel.

Alternatively, in terms of the Boys function
$F_0(x)=\frac{\sqrt{\pi}}{2}\mathrm{erf}(\sqrt{x})/\sqrt{x}$:
\begin{equation}\label{eq:J_ij_M_boys}
  \mathcal{J}_{\vM}^{(ij)}
  = \omega_{\vM}\,\mathcal{S}_{kl}\,
    \frac{2}{\sigma_{ij,s}}\,
    F_0\!\bigl(|\bar{u}_{ij,\vM}|^2/\sigma_{ij,s}^2\bigr),
\end{equation}
which makes the connection to Gaussian basis-set theory explicit.

\paragraph{Neutral electron--electron matrix element.}
\begin{equation}\label{eq:Vee_bare}
  \boxed{V_{kl}^{(ee,\mathrm{bare})}
  = \mathcal{S}_{kl}\sum_{i=1}^{n-1}\sum_{j=i+1}^{n}\sum_{\vM}
    \omega_{\vM}\,
    \frac{\mathrm{erf}\!\bigl(|\bar{u}_{ij,\vM}|/\sigma_{ij,s}\bigr)}
         {|\bar{u}_{ij,\vM}|}.}
\end{equation}

\paragraph{Neutral electron--nuclear matrix element.}
The electron-nuclear interaction involves $|\vrr_i - \vRR_I|$ with the
nucleus at a fixed position $\vRR_I$, so $|\bar{u}_{ij,\vM}|$ is replaced
by $|\bar{\vrr}_{\vM,i}-\vRR_I|$ and $\sigma_{ij,s}\to\sigma_i$:
\begin{equation}\label{eq:VeN_bare}
  \boxed{V_{kl}^{(eN,\mathrm{bare})}
  = -\mathcal{S}_{kl}\sum_{i=1}^n\sum_{I=1}^{N_\mathrm{nuc}} Z_I\sum_{\vM}
    \omega_{\vM}\,
    \frac{\mathrm{erf}\!\bigl(|\bar{\vrr}_{\vM,i}-\vRR_I|/\sigma_i\bigr)}
         {|\bar{\vrr}_{\vM,i}-\vRR_I|}.}
\end{equation}

\paragraph{Nuclear--nuclear interaction.}
$V_{kl}^{(NN)}$ remains a purely classical sum, unchanged from the Ewald case
(Eq.~\eqref{eq:VNN}) with $E_\mathrm{Madelung}$ computed by a convergent direct
lattice sum or simple distance sum for a neutral cell.

\paragraph{Relation to the Ewald formulation.}
Equations~\eqref{eq:Vee_bare} and \eqref{eq:VeN_bare} are equivalent to
taking the $\kappa\to 0$ limit of the Ewald real-space sum restricted to
$\vn=\mathbf{0}$, plus the reciprocal-space sum and self-energy correction all
taken together for a neutral system.  The $\mathrm{erf}/R$ form emerges because
$\mathrm{erfc}(\kappa R)/R\to 1/R$ as $\kappa\to 0$, and the integral
$\int_0^\infty e^{-t^2R^2}dt = \sqrt{\pi}/(2R)$ gives precisely the Boys
$F_0$ result.  For a neutral cell, the $\vn\neq\mathbf{0}$ real-space shells
and the reciprocal-space sum cancel against each other in the $\kappa\to 0$
limit, leaving only the single-image closed-form result.

\subsection{Contact (Dirac delta) operators}
\label{sec:delta}

The operators $\delta^{(3)}(\vrr_i-\vrr_j)$ and $\delta^{(3)}(\vrr_i-\mathbf{S})$
arise in the computation of cusp conditions, electron–electron coalescence
densities, contact hyperfine integrals, and — as shown in
Sec.~\ref{sec:neutral_coulomb} and Appendix~\ref{app:delta} — as the
building blocks of the bare Coulomb matrix element via convolution with
$1/|\mathbf{u}|$.

\subsubsection{Electron--electron contact: $\delta^{(3)}(\vrr_i-\vrr_j)$}

After unfolding and combining the Gaussians, the Fourier representation
of the $\delta$-function converts the $3n$-dimensional integral into an
inverse Fourier transform of a 3D Gaussian in momentum $\mathbf{q}$, with
variance $\sigma_{ij,s}^2/4$ set by the effective pair width
Eq.~\eqref{eq:sigma_ee}.  The result is:
\begin{equation}\label{eq:Dkl_ij_main}
  \boxed{\langle\Phi_k|\delta^{(3)}(\vrr_i-\vrr_j)|\Phi_l\rangle
  = \frac{\mathcal{S}_{kl}}{(\pi\sigma_{ij,s}^2)^{3/2}}
    \sum_{\vM}\omega_{\vM}\,
    e^{-|\bar{u}_{ij,\vM}|^2/\sigma_{ij,s}^2},}
\end{equation}
where $\bar{u}_{ij,\vM}=\bar{\vrr}_{\vM,i}-\bar{\vrr}_{\vM,j}$ is the
mean pair displacement.  The factor $(\pi\sigma_{ij,s}^2)^{-3/2}$ is the
normalization of the 3D Gaussian of width $\sigma_{ij,s}/\sqrt{2}$ that
describes the pair-contact density: a narrower Gaussian (more localized
basis, larger $A_k$) gives a larger contact matrix element.

\subsubsection{Electron at a point: $\delta^{(3)}(\vrr_i-\mathbf{S})$}

For a fixed observation point $\mathbf{S}\in\mathbb{R}^3$ the same
Fourier argument with projector $\vP_i$ yields:
\begin{equation}\label{eq:Dkl_iS_main}
  \boxed{\langle\Phi_k|\delta^{(3)}(\vrr_i-\mathbf{S})|\Phi_l\rangle
  = \frac{\mathcal{S}_{kl}}{(\pi\sigma_i^2)^{3/2}}
    \sum_{\vM}\omega_{\vM}\,
    e^{-|\bar{\vrr}_{\vM,i}-\mathbf{S}|^2/\sigma_i^2},}
\end{equation}
where $\sigma_i^2=(A_{kl}^{-1})_{ii}$ (Eq.~\eqref{eq:sigma_en}) and
$\bar{\vrr}_{\vM,i}$ is the $i$-th 3D block of the combined center
$\bar{\vrr}_\vM$ (Eq.~\eqref{eq:rbar}).  As a function of $\mathbf{S}$
this is a sum of Gaussians centered at $\bar{\vrr}_{\vM,i}$: it equals the
one-electron density matrix element $\gamma_i(\mathbf{S})$ integrated
against $\mathcal{S}_{kl}\omega_\vM$.  Setting $\mathbf{S}=\vRR_I$
gives the contact density at nucleus $I$, which enters the Fermi contact
hyperfine coupling.

\paragraph{Connection to the Coulomb matrix element.}
Both delta results feed directly into the neutral Coulomb matrix elements
of Sec.~\ref{sec:neutral_coulomb} via the convolution identities
\begin{align}
  \mathcal{J}_\vM^{(ij)}&=\int_{\mathbb{R}^3}\frac{\mathcal{D}_\vM^{(ij,\mathbf{u})}}{|\mathbf{u}|}\,d^3\mathbf{u},\label{eq:JfromDelta_ij}\\
  \mathcal{J}_\vM^{(i,I)}&=\int_{\mathbb{R}^3}\frac{\mathcal{D}_\vM^{(i,\mathbf{S})}}{|\mathbf{S}-\vRR_I|}\,d^3\mathbf{S},\label{eq:Jfromelta_iI}
\end{align}
where $\mathcal{D}_\vM^{(ij,\mathbf{u})}$ and $\mathcal{D}_\vM^{(i,\mathbf{S})}$
denote the per-image matrix elements of $\delta^{(3)}(\vrr_i-\vrr_j-\mathbf{u})$
and $\delta^{(3)}(\vrr_i-\mathbf{S})$ respectively.  Evaluating the
convolution with the standard identity
$\int e^{-|\mathbf{t}-\mathbf{R}|^2}/|\mathbf{t}|\,d^3\mathbf{t}
=\pi^{3/2}\mathrm{erf}(|\mathbf{R}|)/|\mathbf{R}|$ recovers
Eqs.~\eqref{eq:Vee_bare} and \eqref{eq:VeN_bare} exactly (see
Appendix~\ref{app:delta} for the full proof).

\subsection{Total Hamiltonian matrix element}

\begin{eqnarray}\label{eq:Hkl}
  H_{kl} &=& T_{kl}
  +V_{kl}^{(ee,\vG)}+V_{kl}^{(ee,\mathrm{real})}
  +V_{kl}^{(eN,\vG)}\nonumber\\
  &+&V_{kl}^{(eN,\mathrm{real})}
  +V_{kl}^{(\mathrm{self})}+V_{kl}^{(NN)}.
  \end{eqnarray}
For a neutral cell one may instead use
\begin{equation}
  H_{kl}=T_{kl}+V_{kl}^{(ee,\mathrm{bare})}
  +V_{kl}^{(eN,\mathrm{bare})}+V_{kl}^{(NN)},
\end{equation}
from Eqs.~\eqref{eq:Vee_bare}--\eqref{eq:VeN_bare}, which is algebraically
simpler and avoids the Ewald parameter $\kappa$ entirely.

\subsection{Antisymmetrization}
\label{sec:antisym}

The expressions above apply to distinguishable-particle basis functions.
For fermions one acts with the antisymmetrizer
$\hat{\mathcal{A}}_\uparrow\hat{\mathcal{A}}_\downarrow$ (restricted to
spin channels), which generates permuted matrix elements.  Under a
permutation $\mathcal{P}\in S_{n_\uparrow}\times S_{n_\downarrow}$, the
ket parameters transform as
\begin{equation}
  A_l^{(\mathcal{P})}=\mathcal{P}A_l\mathcal{P}^T, \qquad
  \vs_l^{(\mathcal{P})}=(\mathcal{P}\otimes I_3)\,\vs_l.
\end{equation}
Since a permuted SCG is again an SCG, the formulas of this section apply
without modification using the permuted parameters.  The antisymmetrized
matrix element is
\begin{equation}
  O_{kl}^{(\mathcal{A})}=
  \sum_{\mathcal{P}\in S_{n_\uparrow}\times S_{n_\downarrow}}
  (-1)^{\mathcal{P}}\,O_{kl}^{(\mathcal{P})},
\end{equation}
where $O_{kl}^{(\mathcal{P})}$ is evaluated with
$A_l\to A_l^{(\mathcal{P})}$, $\vs_l\to\vs_l^{(\mathcal{P})}$.

\section{Specialization: Hydrogen Chain with Two Atoms per Cell}
\label{sec:h2chain}

We specialize the general framework to a one-dimensional chain of hydrogen
atoms with two nuclei and two electrons per primitive cell, using the
neutral-cell Coulomb approach.  All matrix elements are expressed
exclusively in the notation of Sec.~\ref{sec:composite}.

\subsection{Physical setup}

The primitive cell has length $L$ along $\hat{x}$; the $y$ and $z$
directions are treated as open.  Two hydrogen nuclei ($Z_I=1$,
$N_\mathrm{nuc}=2$) sit at
\begin{equation}\label{eq:H2_nuclei}
  \vRR_+ = +a\hat{x}, \qquad \vRR_- = -a\hat{x}, \qquad a = \tfrac{L}{4},
\end{equation}
giving proton--proton spacing $L/2$ and a charge-neutral cell.  There are
$n=2$ electrons with coordinates $(\vrr_1,\vrr_2)\in\mathbb{R}^6$.

\subsection{Basis functions}

The primitive basis function uses the pair-correlation--single-particle
parameterization of Appendix~\ref{app:connection}:
\begin{equation}\label{eq:H2_phi}
  \phi_k(\vrr_1,\vrr_2)
  = \exp\!\left[
      -\frac{1}{2}\sum_{i,j=1}^{2}A^{\mathrm{(pc)}}_{ij}\,\vrr_i\cdot\vrr_j
      -\sum_{i=1}^{2}\beta_i\!\left|\vrr_i - s_i\hat{x}\right|^2
    \right],
\end{equation}
where $A^\mathrm{(pc)}\in\mathbb{R}^{2\times 2}$ is the pair-coupling
matrix, $\beta_i>0$ are single-electron widths, and $s_i\in\mathbb{R}$ are
1D shift parameters.  Completing the square (Appendix~\ref{app:connection})
brings $\phi_k$ to the standard SCG form~\eqref{eq:scg} with
\begin{equation}\label{eq:H2_Ak_sk}
  A_k = \tfrac{1}{2}\bigl(A^\mathrm{(pc)}+2D_\beta\bigr), \qquad
  \vs_k = A_k^{-1}D_\beta\,\mathbf{s}\otimes\hat{x},
\end{equation}
where $D_\beta=\mathrm{diag}(\beta_1,\beta_2)$ and
$\mathbf{s}=(s_1,s_2)^T$.  An overall constant $e^{-\kappa_k}$ (absorbed
into the variational coefficient) arises from the shift:
\begin{equation}\label{eq:H2_kappa}
  \kappa_k = \mathbf{s}^T D_\beta\mathbf{s}
            - \tfrac{1}{2}\vs_k^T A_k\,\vs_k.
\end{equation}
All general formulas apply directly with $A_k$ and $\vs_k$ from
Eq.~\eqref{eq:H2_Ak_sk}.

The periodized basis function is
\begin{equation}\label{eq:H2_Phi}
  \Phi_k(\vrr_1,\vrr_2)
  = \sum_{\mathbf{m}\in\mathbb{Z}^2}
    \phi_k(\vrr_1-m_1 L\hat{x},\,\vrr_2-m_2 L\hat{x}),
\end{equation}
the $n=2$, 1D case of Eq.~\eqref{eq:periodic_basis} with
$\vT_\mathbf{m}=(m_1 L,m_2 L)\hat{x}$.

\subsection{Composite quantities}

For a bra--ket pair $(k,l)$ the composite quantities of
Sec.~\ref{sec:composite} reduce to $2\times 2$ matrices.  Lattice images
are labeled $\mathbf{m}=(m_1,m_2)^T\in\mathbb{Z}^2$, with
\begin{align}\label{eq:H2_TM}
  \vT_\mathbf{m} &= (m_1 L,\,m_2 L)\hat{x},\notag\\
  \vd_\mathbf{m} &= (\vs_k-\vs_l-\vT_\mathbf{m})\big|_x,\notag\\
  \omega_\mathbf{m} &= e^{-\vd_\mathbf{m}^T\widetilde{C}_{kl}\vd_\mathbf{m}}.
\end{align}
The combined Gaussian center has $x$-components
\begin{equation}\label{eq:H2_rbar}
  \bar{\vrr}_{\mathbf{m},i} = \bigl(A_{kl}^{-1}
    [A_k\vs_k + A_l(\vs_l+\vT_\mathbf{m})]\bigr)_i\hat{x}, \quad
    i=1,2,
\end{equation}
and the Gaussian prefactor is $\mathcal{S}_{kl}=\pi^3/(\det A_{kl})^{3/2}$
(Eq.~\eqref{eq:calS} with $n=2$).  The per-image kernel is
\begin{equation}\label{eq:H2_Sab}
  \mathcal{S}_{kl}(\mathbf{m}) \equiv e^{-\kappa_k-\kappa_l}\,
  \omega_\mathbf{m}\,\mathcal{S}_{kl}.
\end{equation}
The effective pair widths (Eqs.~\eqref{eq:sigma_ee}--\eqref{eq:sigma_en})
are
\begin{equation}\label{eq:H2_sigmas}
  \sigma_i^2 = (A_{kl}^{-1})_{ii}, \quad
  \sigma_{12,s}^2
    = (A_{kl}^{-1})_{11}+(A_{kl}^{-1})_{22}-2(A_{kl}^{-1})_{12},
\end{equation}
and the mean pair displacement is
$\bar{u}_{12,\mathbf{m}}=|\bar{\vrr}_{\mathbf{m},1}-\bar{\vrr}_{\mathbf{m},2}|$
(scalar in 1D).

\subsection{Overlap and kinetic energy}

\begin{equation}\label{eq:H2_S}
  S_{kl} = \sum_{\mathbf{m}\in\mathbb{Z}^2}
  \mathcal{S}_{kl}(\mathbf{m}),
\end{equation}
\begin{equation}\label{eq:H2_T}
  T_{kl} = \sum_{\mathbf{m}\in\mathbb{Z}^2}
  \mathcal{S}_{kl}(\mathbf{m})
  \Bigl[3\,\mathrm{Tr}(C_{kl})
       - 2\,\vd_\mathbf{m}^T(C_{kl}^2\otimes I_3)\,\vd_\mathbf{m}\Bigr].
\end{equation}
These are the $n=2$, $\Lambda=I_2$, 1D specializations of
Eqs.~\eqref{eq:Skl} and~\eqref{eq:Tkl}.

\subsection{Neutral-cell Coulomb potential}

Since the cell is charge-neutral the periodic Coulomb sum converges
absolutely when grouped into shells.  The Coulomb operator is decomposed as
\begin{equation}\label{eq:H2_Vnc}
  \hat{V}_\mathrm{nc} = \hat{V}^{(0)}_\mathrm{nc}
  + \sum_{p=1}^{\infty}\hat{V}^{(p)}_\mathrm{nc},
\end{equation}
with the reference-cell ($p=0$) operator
\begin{multline}\label{eq:H2_V0}
  \hat{V}^{(0)}_\mathrm{nc}
  = \frac{1}{|\vrr_1-\vrr_2|}\\
  - \sum_{i=1}^{2}\!\left(\frac{1}{|\vrr_i-a\hat{x}|}
    +\frac{1}{|\vrr_i+a\hat{x}|}\right)
  + \frac{1}{2a},
\end{multline}
and each image shell ($p\geq 1$)
\begin{multline}\label{eq:H2_Vp}
  \hat{V}^{(p)}_\mathrm{nc}
  = \frac{1}{|\vrr_1-\vrr_2-pL\hat{x}|}
    +\frac{1}{|\vrr_1-\vrr_2+pL\hat{x}|}\\
  - \sum_{i=1}^{2}\sum_{\tau=\pm1}\sum_{\sigma=\pm1}
    \frac{1}{|\vrr_i-(\tau pL+\sigma a)\hat{x}|}\\
  + \frac{2}{pL}+\frac{1}{pL-2a}+\frac{1}{pL+2a}.
\end{multline}
Each shell is charge-neutral, ensuring absolute convergence.  With $a=L/4$,
the classical denominators become $1/(2a)=2/L$ and
$1/(pL\pm 2a)=1/((p\pm\tfrac{1}{2})L)$.

\paragraph{Convergence rate of the shell sum.}
The convergence rate of the sum $\sum_{p=1}^\infty \hat{V}^{(p)}_\mathrm{nc}$
is determined by the leading cancellation within each neutral shell.  For
large $p$ the electronic coordinates $\vrr_i$ are negligible compared with
the nuclear separations $pL$ and $pL\pm 2a$.  Expanding each term in
Eq.~\eqref{eq:H2_Vp} in powers of $1/(pL)$, the net charge of the shell
(monopole) vanishes by construction, and the dipole moment of the shell
also vanishes by symmetry (the shell contributes $+2$ from two
electron–electron images and $-4$ from four electron–nuclear images, with
symmetric positions $\pm pL$ and $\pm(pL\pm 2a)$).  The leading surviving
multipole is the quadrupole, so the shell potential felt by an electron at
$\vrr_i\sim O(L)$ from the center falls off as $1/(pL)^3$.  Consequently,
the classical constant contribution (the last line of
Eq.~\eqref{eq:H2_Vp}) also vanishes to order $1/(pL)^2$ and the net
classical correction decays as $1/p^3$.  The matrix elements
$V^{(p)}_{kl}$ therefore satisfy
\begin{equation}\label{eq:shell_convergence}
  \bigl|V^{(p)}_{kl}\bigr| \sim \frac{C_{kl}}{p^3}, \quad p\to\infty,
\end{equation}
where $C_{kl}$ is a basis-dependent constant, giving an absolutely
convergent series whose truncation at shell $p_\mathrm{cut}$ introduces an
error $O(p_\mathrm{cut}^{-2})$.

\paragraph{Ewald split and cutoff condition.}
When using the Ewald decomposition \eqref{eq:ewald} instead, the splitting
parameter $\kappa$ determines the balance between real-space and
reciprocal-space convergence.  The real-space shells $\vn$ satisfy
$\mathrm{erfc}(\kappa|\vn\cdot\vL|)\lesssim\varepsilon$ for
$|\vn\cdot\vL|\gtrsim r_\mathrm{cut}$, giving the real-space cutoff
condition
\begin{equation}\label{eq:rcut}
  r_\mathrm{cut} = \frac{1}{\kappa}\sqrt{\ln(1/\varepsilon)},
\end{equation}
while the reciprocal-space shells satisfy $e^{-G^2/4\kappa^2}\lesssim\varepsilon$
beyond the cutoff
\begin{equation}\label{eq:Gcut}
  G_\mathrm{cut} = 2\kappa\sqrt{\ln(1/\varepsilon)}.
\end{equation}
The product $r_\mathrm{cut}\,G_\mathrm{cut} = 4\ln(1/\varepsilon)$ is
independent of $\kappa$, so the total number of terms scales as
$(r_\mathrm{cut}/L)^3 + (G_\mathrm{cut}L/2\pi)^3 \propto
(\kappa L)^{-3} + (\kappa L)^3$, which is minimized at $\kappa =
\pi^{1/2}/L$ (the classic Ewald optimum).  For the neutral-cell
direct-sum approach the $1/p^3$ decay of Eq.~\eqref{eq:shell_convergence}
replaces both sums with a single series truncated at
$p_\mathrm{cut}\sim(C_{kl}/\varepsilon)^{1/2}$ shells.

The matrix elements are expressed through the per-image Coulomb kernels
\begin{align}
  W_i(x;\mathbf{m}) &\equiv
    \mathcal{S}_{kl}(\mathbf{m})\,
    \frac{\mathrm{erf}\!\bigl(|\bar{\vrr}_{\mathbf{m},i}-x\hat{x}|/\sigma_i\bigr)}
         {|\bar{\vrr}_{\mathbf{m},i}-x\hat{x}|},
  \label{eq:H2_Wi}\\
  W_{12}(d;\mathbf{m}) &\equiv
    \mathcal{S}_{kl}(\mathbf{m})\,
    \frac{\mathrm{erf}\!\bigl(|\bar{u}_{12,\mathbf{m}}-d|/\sigma_{12,s}\bigr)}
         {|\bar{u}_{12,\mathbf{m}}-d|},
  \label{eq:H2_W12}
\end{align}
which are the per-image evaluations of the neutral Coulomb formulas
Eqs.~\eqref{eq:Vee_bare}--\eqref{eq:VeN_bare}.  Equivalently, in terms
of the Boys function $F_0(t)=\frac{\sqrt{\pi}}{2\sqrt{t}}\mathrm{erf}(\sqrt{t})$:
\begin{equation}\label{eq:H2_Boys}
  W_i = \mathcal{S}_{kl}(\mathbf{m})\sqrt{\frac{2}{\pi\sigma_i^2}}\,
  F_0\!\!\left(\frac{|\bar{\vrr}_{\mathbf{m},i}-x\hat{x}|^2}{\sigma_i^2}\right),
\end{equation}
and likewise for $W_{12}$ with $\sigma_i^2\to\sigma_{12,s}^2$.

The matrix element of $\hat{V}^{(0)}_\mathrm{nc}$ is
\begin{multline}\label{eq:H2_V0ab}
  V^{(0)}_{kl} = \sum_{\mathbf{m}\in\mathbb{Z}^2}\left[
    W_{12}(0;\mathbf{m})\right.\\
    \left.-\sum_{i=1}^{2}\bigl(W_i(a;\mathbf{m})+W_i(-a;\mathbf{m})\bigr)
    + \frac{S_{kl}(\mathbf{m})}{2a}\right],
\end{multline}
and for each shell $p\geq 1$:
\begin{multline}\label{eq:H2_Vpab}
  V^{(p)}_{kl} = \sum_{\mathbf{m}\in\mathbb{Z}^2}\Biggl[
    W_{12}(pL;\mathbf{m})+W_{12}(-pL;\mathbf{m})\\
    -\sum_{i=1}^{2}\sum_{\tau=\pm1}\sum_{\sigma=\pm1}
      W_i(\tau pL+\sigma a;\mathbf{m})\\
    +\Bigl(\frac{2}{pL}+\frac{1}{pL-2a}+\frac{1}{pL+2a}\Bigr)
    \mathcal{S}_{kl}(\mathbf{m})\Biggr].
\end{multline}
The total Coulomb matrix element is
$V^\mathrm{nc}_{kl}=V^{(0)}_{kl}+\sum_{p=1}^\infty V^{(p)}_{kl}$.

\subsection{Bloch twist at wave vector $k_\mathrm{B}$}

For a Bloch wave vector $k_\mathrm{B}\in[-\pi/L,\pi/L]$ the periodized
basis carries a phase (Theorem~2, Sec.~\ref{sec:unfolding}):
\begin{equation}\label{eq:H2_Bloch_Phi}
  \Phi_{k,k_\mathrm{B}}(\vrr_1,\vrr_2)
  = \sum_{\mathbf{m}\in\mathbb{Z}^2}
    e^{ik_\mathrm{B}L(m_1+m_2)}\,
    \phi_k(\vrr_1-m_1 L\hat{x},\vrr_2-m_2 L\hat{x}),
\end{equation}
with Bloch phase $e^{i\mathbf{k}_\mathrm{B}\cdot\vT_\mathbf{m}}
=e^{ik_\mathrm{B}L(m_1+m_2)}$.  All real-valued per-image kernels
$\mathcal{S}_{kl}(\mathbf{m})$, $T_{kl}(\mathbf{m})$, $V_{kl}(\mathbf{m})$
are unchanged; the $k_\mathrm{B}$-dependent matrix elements are
\begin{equation}\label{eq:H2_Bloch_O}
  O_{kl}(k_\mathrm{B}) = \sum_{\mathbf{m}\in\mathbb{Z}^2}
  e^{ik_\mathrm{B}L(m_1+m_2)}\,O_{kl}(\mathbf{m}).
\end{equation}
Since the kernels are real and satisfy $O_{lk}(\mathbf{m})=O_{kl}(-\mathbf{m})$,
the matrices $\mathbf{H}(k_\mathrm{B})$ and $\mathbf{S}(k_\mathrm{B})$ are
complex Hermitian.  The generalized eigenproblem
$\mathbf{H}(k_\mathrm{B})\mathbf{c}=E(k_\mathrm{B})\mathbf{S}(k_\mathrm{B})\mathbf{c}$
yields the electronic band structure $E_0(k_\mathrm{B})$.

\subsection{Numerical example}
As an initial test, we set $L$ = 100 and computed the ground-state
energy using 100 basis states optimized via the stochastic variational
method (SVM) at k = 0 (the $\Gamma$ point), obtaining -1.17441 Hartree. This
is in excellent agreement with the precise variational benchmark of
-1.174475 Hartree. Building on this, we applied the SVM to compute
energies across a range of system sizes $L$ on a $k$-point mesh.
Table \ref{tableh2} presents the energy of a periodic hydrogen chain with 2 H
atoms per unit cell, evaluated on a uniform mesh of 33 $k$ points. For
context, Ref. \cite{PhysRevX.7.031059} provides a benchmark study of a finite
(non-periodic) H$_{10}$ chain using a variety of advanced many-body
methods; here we select the auxiliary-field quantum Monte Carlo
(AFQMC) results for comparison, as the other methods yield similar
values. Ref. \cite{PhysRevB.84.245117} reports variational Monte Carlo (VMC) calculations
for periodic hydrogen chains of N = 18, 34, 50, and 66,
employing periodic boundary conditions in all directions with an
elongated supercell in the direction perpendicular to the chain axis.
Our results are in reasonably good agreement with both references.
Some discrepancy with Ref. \cite{PhysRevX.7.031059} is expected, given that it treats a
finite, non-periodic chain. The discrepancy between our results and
those of Ref. \cite{PhysRevB.84.245117} most likely
stems from the limited accuracy of the variational trial function
employed in that work. Specifically, the approach used in Ref.
\cite{PhysRevB.84.245117}
yields an energy of -1.72 Hartree in the large basis set limit, and
this systematic underestimation of the correlation energy is the
probable source of the disagreement with our results.

\begin{table}
\begin{tabular}{|l|l|l|l|}
\hline
L    & This work &  Ref. \cite{PhysRevX.7.031059} & Ref. \cite{PhysRevB.84.245117}\\
\hline
1.0  & -0.40611 &  -0.44284 & -0.41358  \\
1.2  & -0.50331 &  -0.51489 &           \\
1.4  & -0.54213 &  -0.54914 &           \\
1.6  & -0.56499 &  -0.56315 &           \\
1.8  & -0.57283 &  -0.56644 &           \\
2.0  & -0.57253 &  -0.56396 & -0.56284  \\
2.4  & -0.56161 &  -0.55164 &           \\
2.8  & -0.54687 &  -0.53755 &           \\
3.2  & -0.53320 &  -0.53499 &           \\
3.6  & -0.52128 &  -0.51568 &           \\
\hline
\end{tabular}
\caption{Energies (in Hartree) for different bond lengths, $L$ (in Bohr)}
\label{tableh2}
\end{table}
\begin{figure}[h]
\centering
\includegraphics[width=0.5\textwidth]{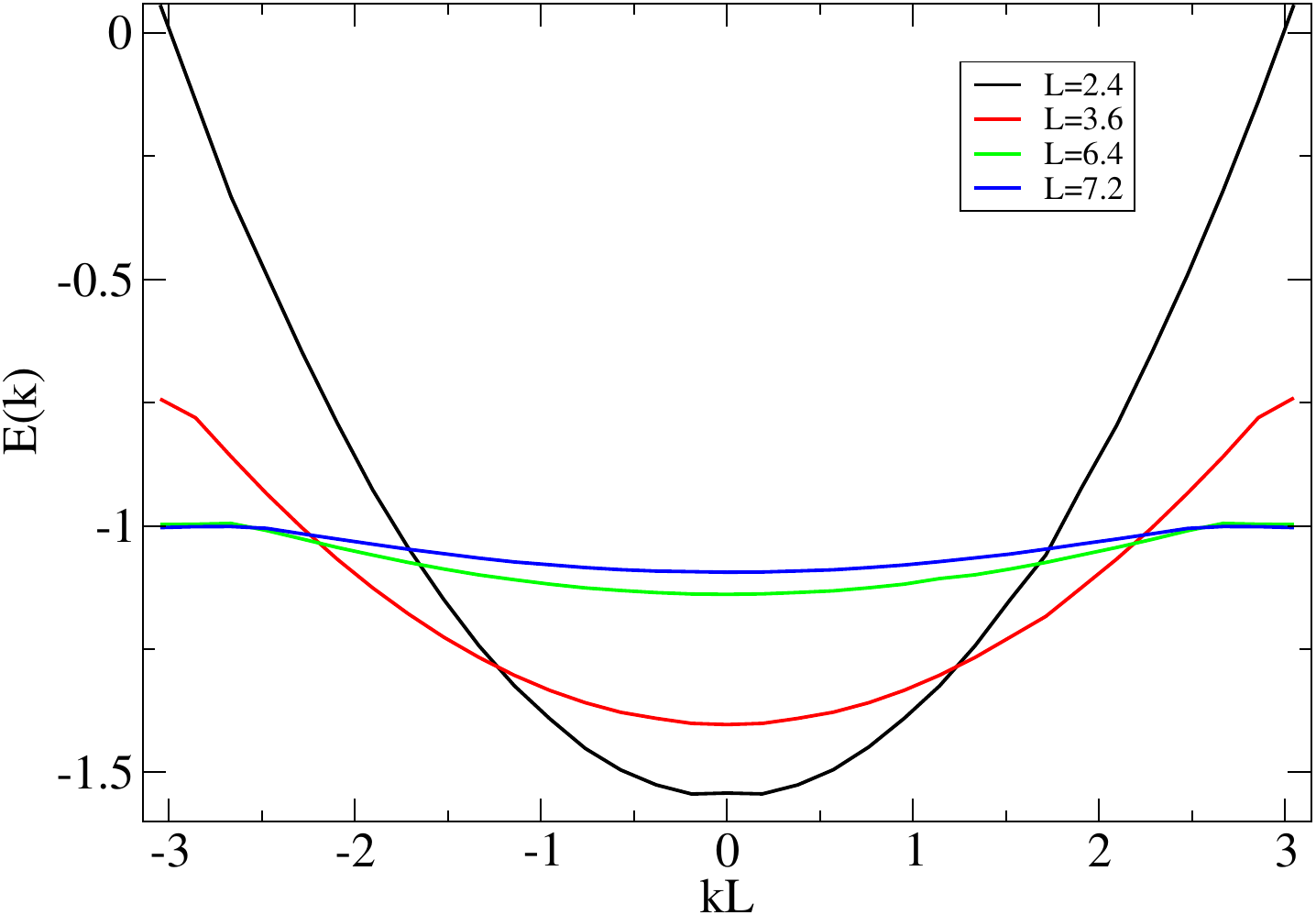}
\caption{Dispersion relation $E(k)$ of the 1D hydrogen chain for
three lattice
constants $L = 2.4$ Bohr (black), $L = 3.6$ Bohr (red), $L=6.4$ (green),
and $L = 7.2$ Bohr (blue). The horizontal axis is
$k\cdot L$; the vertical axis
is the band energy in Hartree.}
\label{fig:bands}
\end{figure}

Fig.~\ref{fig:bands} shows the single-band
dispersion relation for four
lattice constants. All bands are symmetric about
$k = 0$ (the $\Gamma$ point),
as required by time-reversal symmetry.

To assess how well a simple tight-binding picture
captures the correlated
electronic structure, we fitted our explicitly
correlated Gaussian results
to a nearest-neighbor tight-binding model. For a
one-dimensional hydrogen
chain with periodic boundary conditions, this
model gives the band dispersion
\begin{equation}
E(k) = \epsilon_0 + 2t \cos(kL),
\end{equation}
where $L$ is the lattice constant, $\epsilon_0$
is the on-site energy, and
$t$ is the nearest-neighbor hopping parameter.
Within this model, the
bandwidth --- defined as the difference between
the maximum and minimum band
energy --- is simply
\begin{equation}
W = 4|t|.
\end{equation}
A large bandwidth reflects strong hopping and
significant orbital overlap
between adjacent hydrogen atoms, while a small
bandwidth indicates more
localized states and a correspondingly flat dispersion.

The nearest-neighbor tight-binding fits are
summarized in
Table~\ref{nnfit}.

\begin{table}[h]
\centering
\caption{Nearest-neighbor tight-binding
parameters obtained by fitting to
the explicitly correlated Gaussian band
structure of the one-dimensional
hydrogen chain. All energies are in Hartree.}
\begin{tabular}{c c c c c c}
\hline
\(L\)  &  \(W\) &
\(\epsilon_0\) &
\(t\) & RMS error & Max error \\
\hline
2.4 & 1.601  & -0.9822 & -0.3449 & 0.1402  &
0.3527 \\
3.6 & 0.6626 & -1.1546 & -0.1497 & 0.0546  &
0.1162 \\
6.4 & 0.1439 & -1.0735 & -0.0368 & 0.00732 &
0.0134 \\
7.2 & 0.0927 & -1.0494 & -0.0243 & 0.00346 &
0.00643 \\
\hline
\end{tabular}
\label{nnfit}
\end{table}

The negative values of $t$ confirm that the
band minimum occurs at
$kL = 0$, consistent with a bonding ground
state. The bandwidth decreases
rapidly with increasing lattice constant,
reflecting the expected exponential
suppression of orbital overlap as the hydrogen
atoms are moved further apart.

The quality of the fit varies strongly with
cell size. For the two largest
cells ($L = 6.4$ and $7.2$ Bohr), the RMS
errors are small, indicating that
the nearest-neighbor model provides an
accurate description of the band
structure. For the smaller cells ($L = 2.4$
and $3.6$ Bohr), however, the
errors are considerably larger. In this
regime, the explicitly correlated
Gaussians have significant amplitude on atoms
beyond the nearest neighbor,
coupling hydrogen atoms across multiple unit
cells. A nearest-neighbor
tight-binding model is therefore insufficient
to capture the full dispersion,
and longer-range hopping terms would be
required for a faithful representation.

\section{Summary and Outlook}
\label{sec:summary}

This paper develops a complete variational framework for computing the
electronic structure of periodic solids using a basis of shifted
correlated Gaussians (SCGs).  The central achievement is the derivation
of closed-form expressions for every matrix element that enters the
generalized eigenvalue problem $\mathbf{H}\mathbf{c}=E\mathbf{S}\mathbf{c}$,
including the overlap, kinetic energy, electron--electron and
electron--nuclear Coulomb interactions, nuclear--nuclear (Madelung)
energy, Dirac delta contact operators, and their Bloch-phase
generalizations for band-structure calculations at arbitrary wave
vector $\mathbf{k}_\mathrm{B}$.

The key technical advance that makes the derivation tractable is the
\emph{generalized unfolding theorem} (Theorem~1 and its Bloch
extension, Theorem~2).  When two periodized basis functions
$\Phi_k(\vrr)=\sum_\vM\phi_k(\vrr-\vT_\vM)$ are inserted into a matrix
element over the simulation cell $\Omega^n$, a double lattice sum
arises.  The unfolding theorem collapses this double sum into a single
sum over image offsets $\vM\in\mathbb{Z}^{3n}$, simultaneously
promoting the cell integral to an all-space integral over the
non-periodized Gaussians $\phi_k$, $\phi_l$.  Because individual
Gaussians are not lattice-periodic, their all-space integrals admit
analytic evaluation in closed form, which would not be possible within
the finite simulation cell.  The resulting image sum is weighted by
$\omega_\vM=e^{-\vd_\vM^T\widetilde{C}_{kl}\vd_\vM}$, which decays
exponentially in $|\vT_\vM|^2$ due to the positive definiteness of
$\widetilde{C}_{kl}=C_{kl}\otimes I_3$, guaranteeing absolute
convergence and reducing the infinite sum to a finite shell of images
in practice.

The long-range Coulomb interaction under periodic boundary conditions
is handled via three independent and mutually consistent methods.
The \emph{Ewald decomposition} splits the conditionally convergent
$1/r$ lattice sum into a short-range complementary-error-function
part evaluated in real space and a smooth long-range part evaluated in
reciprocal space; individual divergences in the electron--electron,
electron--nuclear, and self-energy terms cancel exactly in the
physical charge-neutral combination, yielding a finite result
independent of the splitting parameter $\kappa$.  For charge-neutral
cells, a simpler \emph{direct neutral-shell sum} is available: the
bare $1/r$ lattice sum grouped into neutral shells is absolutely
convergent, and the matrix element reduces analytically to a screened
Coulomb potential $\mathrm{erf}(R/\sigma_{ij})/R$ with no free
parameter.  A third route, the \emph{Dirac delta convolution method},
expresses the Coulomb matrix element as a weighted integral of the
pair-contact density $\langle\Phi_k|\delta^{(3)}(\vrr_i-\vrr_j-\mathbf{u})|\Phi_l\rangle$
and recovers the neutral-shell result exactly, confirming the
equivalence of all three approaches.  The delta matrix elements carry
independent physical content, entering the electron--electron cusp
condition, the pair-correlation function $g(r)$, and the Fermi
contact hyperfine coupling.

A unifying structural feature is that every matrix element — overlap,
kinetic energy, Ewald reciprocal-space terms, real-space Coulomb terms,
and contact operators — factors into the same form: a Gaussian
prefactor $\mathcal{S}_{kl}=\pi^{3n/2}/(\det A_{kl})^{3/2}$, which
encodes all quantum-mechanical content and is computed once per basis
pair $(k,l)$, multiplied by an image sum weighted by $\omega_\vM$.
This unification means a single numerical infrastructure suffices for
all operator classes.  The Bloch-phase extension at wave vector
$\mathbf{k}_\mathrm{B}$ requires no new integrals: it is simply a
phase-weighted Fourier transform of the same per-image integrals
already computed at the $\Gamma$-point,
$O_{kl}(\mathbf{k}_\mathrm{B})=\sum_\vM e^{i\mathbf{k}_\mathrm{B}\cdot\vT_\vM}O_{kl}(\vM)$,
making band-structure calculations across the Brillouin zone
computationally straightforward.

The formalism is validated through application to an infinite
one-dimensional hydrogen chain with two atoms per primitive cell.  The
ground-state energy per atom computed in the thermodynamic limit agrees
with finite-chain results extrapolated by state-of-the-art coupled-cluster
and quantum Monte Carlo methods, demonstrating both the correctness of
the matrix element expressions and the convergence of the ECG variational
energy with basis size.

Table~\ref{tab:summary} collects all matrix elements with the shorthand
$\mathcal{S}_{kl}$ (Eq.~\ref{eq:calS}) and $\omega_{\vM}$
(Eq.~\ref{eq:omegaM}).
\begin{table*}[htb]
\caption{Summary of matrix elements.  Shorthand:
$\mathcal{S}_{kl}=\pi^{3n/2}/(\det A_{kl})^{3/2}$,
$\omega_{\vM}=e^{-\vd_{\vM}^T\widetilde{C}_{kl}\vd_{\vM}}$,
$R_{ij,\vM}=|\bar{\vrr}_{\vM,i}-\bar{\vrr}_{\vM,j}|$,
$R_{iI,\vM}=|\bar{\vrr}_{\vM,i}-\vRR_I|$,
$R_{iS,\vM}=|\bar{\vrr}_{\vM,i}-\mathbf{S}|$.
All sums over $\vM$ run over $\mathbb{Z}^{3n}$;
$i,j$ run from 1 to $n$; $I$ from 1 to $N_\mathrm{nuc}$.}
\label{tab:summary}
\begin{center}
\renewcommand{\arraystretch}{1.5}
\begin{tabular}{p{10cm}r}
\hline\hline
Matrix element & Eq. \\
\hline
$\displaystyle S_{kl}=\mathcal{S}_{kl}\sum_{\vM}\omega_{\vM}$
  & \eqref{eq:Skl}\\[4pt]
$\displaystyle T_{kl}=\frac{1}{2}\,\mathcal{S}_{kl}\sum_{\vM}\omega_{\vM}
  \bigl[6\,\Tr(\Lambda C_{kl})
  -4\,\vd_{\vM}^T(B_{kl}^{(\Lambda)}\otimes I_3)\vd_{\vM}\bigr]$
  & \eqref{eq:Tkl}\\
\multicolumn{2}{l}{\textit{Contact operators}}\\
$\displaystyle\langle\Phi_k|\delta^{(3)}(\vrr_i-\vrr_j)|\Phi_l\rangle
  =\frac{\mathcal{S}_{kl}}{(\pi\sigma_{ij,s}^2)^{3/2}}
  \sum_\vM\omega_\vM\,e^{-R_{ij,\vM}^2/\sigma_{ij,s}^2}$
  & \eqref{eq:Dkl_ij_main}\\
$\displaystyle\langle\Phi_k|\delta^{(3)}(\vrr_i-\mathbf{S})|\Phi_l\rangle
  =\frac{\mathcal{S}_{kl}}{(\pi\sigma_i^2)^{3/2}}
  \sum_\vM\omega_\vM\,e^{-R_{iS,\vM}^2/\sigma_i^2}$
  & \eqref{eq:Dkl_iS_main}\\
\multicolumn{2}{l}{\textit{Ewald (periodic, all charge configurations)}}\\
$V_{kl}^{(ee,\vG)}$: reciprocal-space $ee$, see Eq.~\eqref{eq:Vee_G}
  & \eqref{eq:Vee_G}\\
$V_{kl}^{(ee,\mathrm{real})}$: real-space $ee$, see Eq.~\eqref{eq:Vee_real}
  & \eqref{eq:Vee_real}\\
$V_{kl}^{(eN,\vG)}$: reciprocal-space $eN$, see Eq.~\eqref{eq:VeN_G}
  & \eqref{eq:VeN_G}\\
$V_{kl}^{(eN,\mathrm{real})}$: real-space $eN$, see Eq.~\eqref{eq:VeN_real}
  & \eqref{eq:VeN_real}\\
$V_{kl}^{(\mathrm{self})}$: Ewald self-energy, see Eq.~\eqref{eq:Vself}
  & \eqref{eq:Vself}\\
$\displaystyle V_{kl}^{(NN)}=E_\mathrm{Madelung}\cdot S_{kl}$
  & \eqref{eq:VNN}\\
\multicolumn{2}{l}{\textit{Direct (neutral cell, no Ewald parameter $\kappa$)}}\\
$\displaystyle V_{kl}^{(ee,\mathrm{bare})}
  =\mathcal{S}_{kl}\sum_{i<j}^{n}\sum_\vM\omega_\vM\,
  \frac{\mathrm{erf}(R_{ij,\vM}/\sigma_{ij,s})}{R_{ij,\vM}}$
  & \eqref{eq:Vee_bare}\\
$\displaystyle V_{kl}^{(eN,\mathrm{bare})}
  =-\mathcal{S}_{kl}\sum_{i=1}^{n}\sum_{I=1}^{N_\mathrm{nuc}}Z_I
  \sum_\vM\omega_\vM\,
  \frac{\mathrm{erf}(R_{iI,\vM}/\sigma_i)}{R_{iI,\vM}}$
  & \eqref{eq:VeN_bare}\\
\hline\hline
\end{tabular}
\end{center}
\end{table*}
Note the structural parallel between the contact and bare
Coulomb rows: the latter are the convolution of the former with
$1/|\mathbf{u}|$, replacing $e^{-R^2/\sigma^2}/(\pi\sigma^2)^{3/2}$
with the screened potential $\mathrm{erf}(R/\sigma)/R$.

The ECG framework derived here is applicable to a broad class of periodic
systems, with the practical constraint being the number of explicitly
treated electrons per primitive cell $n$: ECG basis sizes scale
exponentially with $n$, so the approach is most powerful for systems with
few valence electrons, especially where electron correlation beyond
mean-field theory is physically essential.

The most natural starting points are model periodic systems that also serve
as benchmarks.  \textit{Hydrogen chains and crystals} are the paradigmatic
correlated-electron problem in one, two, and three dimensions.  The 1D
equispaced hydrogen chain (one $1s$ electron per site, one electron per
unit cell) exhibits a Mott–Hubbard metal–insulator transition as a
function of lattice spacing and has been studied with coupled-cluster and
QMC methods; the present framework treats the long-range Coulomb
interaction exactly without a model Hamiltonian.  The 2D square hydrogen
lattice (one atom and one electron per primitive cell) extends this to a
genuinely two-dimensional correlated metal–insulator problem.  The 3D
simple-cubic atomic hydrogen crystal (again one electron per cell) is the
simplest three-dimensional case where accurate treatment of Coulomb
correlations is required to describe the metal–insulator transition
predicted around 400 GPa \cite{PhysRevLett.124.116401}.  
In all three geometries the charge-neutral
direct-sum formulas of Sec.~\ref{sec:neutral_coulomb} apply directly,
providing a parameter-free alternative to Ewald for the Coulomb matrix
elements.

Among simple metals, \textit{solid lithium} (body-centered cubic, BCC,
lattice constant $a = 3.51$\,\AA) and \textit{solid sodium} (BCC,
$a = 4.29$\,\AA) are the natural first targets.  Each has one valence
electron per atom and one atom per primitive cell, giving $n=1$ explicit
electron when the [He] or [Ne] core is represented by a pseudopotential.
Lithium and sodium are nearly-free-electron metals at ambient pressure
whose correlation energy is nonetheless important for accurate cohesive
energies and bulk moduli, as demonstrated by QMC pseudopotential
calculations~\cite{PhysRevB.101.165125}.  

\textit{Solid aluminum} (face-centered cubic, FCC, $a = 4.05$\,\AA)
adds one further level of complexity: the primitive cell contains one
atom with three valence electrons ($3s^23p^1$), giving $n=3$ when the
[Ne] core is frozen into a pseudopotential.  With three electrons per cell
the basis remains tractable for ECG methods, while the $s$–$p$ mixed
valence and the presence of a Fermi surface make the band structure and
correlation energy richer than in the alkali metals.  Coupled-cluster
calculations for FCC aluminum using a primitive two-atom cell have
recently been reported; the ECG approach would provide complementary
all-order electron-correlation results within a single cell.

\textit{Graphene} is a prototypical 2D periodic system: its hexagonal
primitive cell contains two inequivalent carbon atoms separated by
$1.42$\,\AA\ (lattice constant $2.46$\,\AA).  With a pseudopotential
freezing the $1s^2$ core of each carbon, one retains two valence electrons
per carbon ($2s^22p^2$ reduced to the frontier $\pi$ electron per atom),
giving $n=2$ or $n=4$ depending on whether only the $\pi$ bands or all
valence electrons are treated explicitly.  Graphene's linear Dirac
dispersion at the $K$ and $K'$ points of the hexagonal Brillouin zone and
its zero gap make it a stringent test of $k$-point convergence; the
Bloch-phase extension (Theorem~2) is essential, requiring matrix element
evaluation at a dense mesh of $\mathbf{k}_\mathrm{B}$ points to resolve
the Dirac cones.

More generally, any solid with few valence electrons and a small primitive
cell — alkali halides (one ion pair per cell, e.g.\ LiH with $n=2$
valence electrons), boron nitride monolayer (two atoms, $n=4$), or
hydrogen-rich superconductors under pressure — falls within the scope of
the present formalism.  The pseudopotential representation of chemically
inert core electrons reduces $n$ to the physically active valence count,
making ECG an attractive high-accuracy complement to plane-wave
density-functional methods for these systems.

\section*{Acknowledgments}
This work was supported by the National Science Foundation
(NSF) under Grant No. DMR-2217759. 
\section*{Data Availability Statement}
The data that support the findings of this study are available
from the corresponding author upon reasonable request.


\appendix

\section{Proof of the Unfolding Theorem}
\label{app:unfolding}

\subsection{Setup and notation}

Let $\Omega^n\subset\mathbb{R}^{3n}$ be the fundamental domain (simulation
cell for all $n$ electrons).  The lattice of composite translations
$\mathcal{L}^n=\{\vT_{\vM}:\vM\in\mathbb{Z}^{3n}\}$ tiles $\mathbb{R}^{3n}$
exactly, i.e.\
$\mathbb{R}^{3n}=\bigsqcup_{\vM\in\mathbb{Z}^{3n}}(\Omega^n+\vT_{\vM})$.
The periodized basis functions are $\Phi_k(\vrr)=\sum_{\vM}\phi_k(\vrr-\vT_{\vM})$.

\subsection{Proof of Theorem 1}

The matrix element of $\hat{O}$ in the periodized basis is, by definition,
\begin{equation}
  O_{kl}=\int_{\Omega^n}\Phi_k^*(\vrr)\,\hat{O}\,\Phi_l(\vrr)\,d\vrr.
\end{equation}
Expanding both sums:
\begin{equation}
  O_{kl}=\sum_{\vM_k,\vM_l\in\mathbb{Z}^{3n}}
  \int_{\Omega^n}
  \phi_k^*(\vrr-\vT_{\vM_k})\,\hat{O}\,\phi_l(\vrr-\vT_{\vM_l})\,d\vrr.
\end{equation}

\paragraph{Step 1: Shift the integration variable.}
For each fixed $\vM_k$, apply the change of variables $\vrr\mapsto\vrr+\vT_{\vM_k}$:
\begin{equation}
  O_{kl}=\sum_{\vM_k,\vM_l}
  \int_{\Omega^n+\vT_{\vM_k}}
  \phi_k^*(\vrr)\,\bigl[\hat{O}\,\phi_l\bigr](\vrr-\vT_{\vM_l-\vM_k})\,d\vrr.
\end{equation}
Here we used the lattice-periodicity condition
Eq.~\eqref{eq:kernel_periodic}: the kernel of $\hat{O}$ satisfies
$K(\vrr+\vT_{\vM_k},\vrr'+\vT_{\vM_k})=K(\vrr,\vrr')$, so acting with $\hat{O}$
and then shifting is the same as first shifting the argument of $\phi_l$.
Concretely, for any function $f$:
\begin{equation}
  \bigl[\hat{O}\,f(\cdot-\vT_{\vM_l})\bigr](\vrr+\vT_{\vM_k})
  =\bigl[\hat{O}\,f(\cdot-\vT_{\vM_l-\vM_k})\bigr](\vrr).
\end{equation}

\paragraph{Step 2: Promote the cell integral to all space.}
Since $\{\Omega^n+\vT_{\vM_k}\}_{\vM_k\in\mathbb{Z}^{3n}}$ is a disjoint
partition of $\mathbb{R}^{3n}$, summing over $\vM_k$ converts the integral
over each translated cell into a single integral over all of $\mathbb{R}^{3n}$:
\begin{equation}
  O_{kl}=\sum_{\vM_l-\vM_k=:\vM}\left(\sum_{\vM_k}1\right)
  \int_{\mathbb{R}^{3n}}
  \phi_k^*(\vrr)\,\hat{O}\,\phi_l(\vrr-\vT_{\vM})\,d\vrr.
\end{equation}
The inner sum $\sum_{\vM_k}$ is simply the number of times a fixed difference
$\vM=\vM_l-\vM_k$ is realized, which is \emph{exactly once} for each
$\vM\in\mathbb{Z}^{3n}$.  This yields Eq.~\eqref{eq:unfolding}. \hfill$\square$

\paragraph{Remark on convergence.}
The interchange of sum and integral is justified whenever
$\phi_k(\vrr)\,(\hat{O}\phi_l)(\vrr-\vT_{\vM})$ is absolutely summable over
$\vM$ in $L^1(\mathbb{R}^{3n})$.  For Gaussian basis functions, the integrand
decays as $e^{-\vd_{\vM}^T\widetilde{C}_{kl}\vd_{\vM}}$ (exponentially in
$|\vM|^2$), so this is always satisfied.

\subsection{Proof of Theorem 2 (Bloch generalization)}

The Bloch matrix element is
\begin{widetext}
\begin{equation}
  O_{kl}^{(\mathbf{k}_\mathrm{B})}
  =\sum_{\vM_k,\vM_l}
   e^{-i\mathbf{k}_\mathrm{B}\cdot\vT_{\vM_k}}
   e^{+i\mathbf{k}_\mathrm{B}\cdot\vT_{\vM_l}}
   \int_{\Omega^n}
   \phi_k^*(\vrr-\vT_{\vM_k})\,\hat{O}\,\phi_l(\vrr-\vT_{\vM_l})\,d\vrr.
\end{equation}
\end{widetext}
Applying Steps 1--2 of the Theorem 1 proof, the $\vM_k$-sum again promotes
the integral to $\mathbb{R}^{3n}$, leaving the phase
$e^{i\mathbf{k}_\mathrm{B}\cdot(\vT_{\vM_l}-\vT_{\vM_k})}
=e^{i\mathbf{k}_\mathrm{B}\cdot\vT_{\vM}}$ (with $\vM=\vM_l-\vM_k$)
attached to each term, giving Eq.~\eqref{eq:unfolding_bloch}. \hfill$\square$

\subsection{Classification of operator types and resulting image sums}

Table~\ref{tab:operators} summarizes how the unfolded integrand
$\mathcal{I}_\vM\equiv\int_{\mathbb{R}^{3n}}\phi_k\,\hat{O}\,\phi_l(\cdot-\vT_\vM)\,d\vrr$
depends on the operator class.

\begin{table}[h]
\caption{Unfolded integrand structure by operator class.  The image weight is
$\omega_\vM=e^{-\vd_\vM^T\widetilde{C}_{kl}\vd_\vM}$ and $\mathcal{S}_{kl}$ is
defined in Eq.~\eqref{eq:calS}.  ``Poly$_\vM$'' denotes a polynomial in $\vd_\vM$
arising from Gaussian moments.}
\label{tab:operators}
\begin{center}
\begin{tabular}{p{3.2cm}p{4.6cm}}
\hline\hline
Operator class & $\mathcal{I}_\vM$ structure \\
\hline
Identity (overlap) & $\mathcal{S}_{kl}\,\omega_\vM$ \\[3pt]
Differential $\hat{D}(\nabla)$ & $\mathcal{S}_{kl}\,\omega_\vM\cdot\mathrm{Poly}_\vM(\vd_\vM)$ \\[3pt]
Local periodic $V(\vrr)$ & $\mathcal{S}_{kl}\,\omega_\vM\cdot\hat{V}$-dependent \\[3pt]
Ewald (real space) & $\mathcal{S}_{kl}\sum_\vn(\mathrm{function\ of\ }\vd_\vM,\vn)$ \\[3pt]
Ewald (recip.\ space) & $\mathcal{S}_{kl}\,\omega_\vM\sum_\vG(\mathrm{phase\ factor})$ \\[3pt]
Bare $1/r$ (neutral cell) & $\mathcal{S}_{kl}\,\omega_\vM\cdot\pi^{3/2}\sigma^{-2}\mathrm{erf}(R/\sigma)/R$ \\[3pt]
Bloch ($\mathbf{k}_\mathrm{B}\neq\mathbf{0}$) & above $\times\,e^{i\mathbf{k}_\mathrm{B}\cdot\vT_\vM}$ \\
\hline\hline
\end{tabular}
\end{center}
\end{table}

\noindent In every case $O_{kl}=\sum_\vM\mathcal{I}_\vM$, and the Gaussian
decay of $\omega_\vM$ (or its analogue after augmentation) ensures convergence.

\section{Derivation of the Overlap Matrix Element}
\label{app:overlap}

\subsection{Combining the two Gaussians}

Theorem~1 with $\hat{O}=1$ reduces the overlap to
$S_{kl}=\sum_\vM\int_{\mathbb{R}^{3n}}\phi_k(\vrr)\phi_l(\vrr-\vT_\vM)\,d\vrr$.
The product $\phi_k(\vrr)\phi_l(\vrr-\vT_{\vM})$ is
\begin{multline}
  \phi_k(\vrr)\phi_l(\vrr-\vT_{\vM})\\
  =\exp\!\bigl[-(\vrr-\vs_k)^T\At_k(\vrr-\vs_k)
              -(\vrr-\vs_l-\vT_{\vM})^T\At_l(\vrr-\vs_l-\vT_{\vM})\bigr].
\end{multline}
Expanding and completing the square in $\vrr$ yields
\begin{equation}
  \phi_k(\vrr)\phi_l(\vrr-\vT_{\vM})=
  \omega_{\vM}\,e^{-(\vrr-\bar{\vrr}_{\vM})^T\At_{kl}(\vrr-\bar{\vrr}_{\vM})},
\end{equation}
where $\omega_{\vM}=e^{-\vd_{\vM}^T\widetilde{C}_{kl}\vd_{\vM}}$ is
verified by direct algebra using $C_{kl}=A_kA_{kl}^{-1}A_l$ (see below).

\paragraph{Verification of $\omega_{\vM}$.}
The residual after completing the square is
\begin{align}
  Q_{\vM}&=\vs_k^T\At_k\vs_k+(\vs_l+\vT_{\vM})^T\At_l(\vs_l+\vT_{\vM})
            -\bar{\vrr}_{\vM}^T\At_{kl}\bar{\vrr}_{\vM}.
\end{align}
Substituting $\bar{\vrr}_{\vM}=\At_{kl}^{-1}[\At_k\vs_k+\At_l(\vs_l+\vT_{\vM})]$
and using the identity
$A_kA_{kl}^{-1}A_k+A_kA_{kl}^{-1}A_l=A_k$:
\begin{equation}
  Q_{\vM}=\vd_{\vM}^T(C_{kl}\otimes I_3)\vd_{\vM}=\vd_{\vM}^T\widetilde{C}_{kl}\vd_{\vM}.
\end{equation}

\subsection{Gaussian integral}

Integrating over $\mathbb{R}^{3n}$ with $N=3n$:
\begin{equation}
  \int_{\mathbb{R}^{3n}}e^{-(\vrr-\bar{\vrr})^T\At_{kl}(\vrr-\bar{\vrr})}d\vrr
  =\frac{\pi^{3n/2}}{(\det\At_{kl})^{1/2}}
  =\frac{\pi^{3n/2}}{(\det A_{kl})^{3/2}},
\end{equation}
giving Eq.~\eqref{eq:Skl}.

\section{Derivation of the Kinetic Energy Matrix Element}
\label{app:kinetic}

\subsection{Laplacian of a shifted Gaussian}

Let $\vrr'=\vrr-\vs_l-\vT_{\vM}$.  The mass-weighted Laplacian acting on
$\phi_l(\vrr-\vT_{\vM})=e^{-\vrr'^T\At_l\vrr'}$ gives
\begin{equation}\label{eq:lap_gauss}
  \sum_{i}\frac{1}{m_i}\nabla_i^2\,\phi_l
  =\Bigl[4\,\vrr'^T\At_l\widetilde{\Lambda}\At_l\vrr'
         -2\,\Tr(\widetilde{\Lambda}\At_l)\Bigr]\phi_l,
\end{equation}
where $\widetilde{\Lambda}=\Lambda\otimes I_3$ and we used
$\Tr(\widetilde{\Lambda}\At_l)=\Tr(\Lambda A_l)\cdot\Tr(I_3)=3\Tr(\Lambda A_l)$.

\subsection{Constant (trace) contribution}

After applying Theorem~1 (Appendix~\ref{app:unfolding}, Corollary~1b) to the
differential operator $\hat{O}=-\frac{1}{2}\sum_i m_i^{-1}\nabla_i^2$, the
trace term gives a multiple of the overlap integral.
Using the identity $A_l=A_{kl}-A_k$ and cyclic trace properties:
\begin{equation}
  6\Tr(\Lambda A_l)-6\Tr(A_l\Lambda A_lA_{kl}^{-1})=6\Tr(\Lambda C_{kl}).
\end{equation}
This simplification is established by writing
$A_lA_{kl}^{-1}=I-A_kA_{kl}^{-1}$ and using $A_lA_{kl}^{-1}A_k=C_{kl}$.

\subsection{Quadratic contribution}

The quadratic term in Eq.~\eqref{eq:lap_gauss} requires the Gaussian second
moment.  With $\vu=\vrr-\bar{\vrr}_{\vM}$ and
$\bm{\delta}_{\vM}=\bar{\vrr}_{\vM}-\vs_l-\vT_{\vM}=(A_{kl}^{-1}A_k\otimes I_3)\vd_{\vM}$,
\begin{align}
  \int\vrr'^T\At_l\widetilde{\Lambda}\At_l\vrr'\,
    e^{-\vu^T\At_{kl}\vu}d\vu
  &=\frac{\pi^{3n/2}}{(\det A_{kl})^{3/2}}
    \Bigl[\tfrac{3}{2}\Tr(A_l\Lambda A_lA_{kl}^{-1})
    \notag\\&\qquad\qquad
          +\bm{\delta}_{\vM}^T\At_l\widetilde{\Lambda}\At_l\bm{\delta}_{\vM}\Bigr].
\end{align}

Substituting $\bm{\delta}_{\vM}=(A_{kl}^{-1}A_k\otimes I_3)\vd_{\vM}$:
\begin{equation}
  \bm{\delta}_{\vM}^T\At_l\widetilde{\Lambda}\At_l\bm{\delta}_{\vM}
  =\vd_{\vM}^T(A_kA_{kl}^{-1}A_l\Lambda A_lA_{kl}^{-1}A_k\otimes I_3)\vd_{\vM}.
\end{equation}
Expanding $A_l=A_{kl}-A_k$ in both factors and simplifying using $C_{kl}=A_kA_{kl}^{-1}A_l$
yields
\begin{equation}
  A_kA_{kl}^{-1}A_l\Lambda A_lA_{kl}^{-1}A_k = C_{kl}\Lambda C_{kl}
  \equiv B_{kl}^{(\Lambda)}.
\end{equation}

\subsection{Combining and final simplification}

Assembling the constant and quadratic pieces and applying the trace
simplification established above gives Eq.~\eqref{eq:Tkl}.

\section{Derivation of the Reciprocal-Space Coulomb Matrix Elements}
\label{app:recip_coulomb}

\subsection{Fourier-modulated Gaussian overlap}

We need the integral
\begin{equation}
  \mathcal{I}_{\vM}(\vk)
  =\int_{\mathbb{R}^{3n}}\phi_k(\vrr)\,e^{i\vk^T\vrr}\,\phi_l(\vrr-\vT_{\vM})\,d\vrr.
\end{equation}
Using the product formula of Appendix~\ref{app:overlap} and completing the
square with the linear phase,
\begin{equation}
  \mathcal{I}_{\vM}(\vk)
  =\omega_{\vM}\,e^{i\vk^T\bar{\vrr}_{\vM}}\,\mathcal{S}_{kl}\,
   e^{-\vk^T\At_{kl}^{-1}\vk/4},
\end{equation}
which follows from the standard identity
$\int_{\mathbb{R}^N}e^{-\vx^TM\vx+i\vk^T\vx}d\vx
=\pi^{N/2}(\det M)^{-1/2}e^{-\vk^TM^{-1}\vk/4}$ ($M\succ 0$).

\subsection{Electron--electron reciprocal space}

Setting $\vk=\vP_{ij}\vG=(\ve_i-\ve_j)\otimes I_3\cdot\vG$:
\begin{equation}
  \vk^T\At_{kl}^{-1}\vk=\sigma_{ij,s}^2\,G^2, \qquad
  \vk^T\bar{\vrr}_{\vM}=\vG\cdot(\bar{\vrr}_{\vM,i}-\bar{\vrr}_{\vM,j}).
\end{equation}
Summing the reciprocal-space Ewald term from Eq.~\eqref{eq:ewald} over
all pairs and images yields Eq.~\eqref{eq:Vee_G}.

\subsection{Electron--nuclear reciprocal space}

Setting $\vk=\vP_i\vG$ and including the nuclear phase $e^{-i\vG\cdot\vRR_I}$
gives $\vk^T\At_{kl}^{-1}\vk=\sigma_i^2 G^2$ and leads to Eq.~\eqref{eq:VeN_G}.

\section{Derivation of the Real-Space Coulomb Matrix Elements}
\label{app:real_coulomb}

\subsection{Integral representation of erfc$/r$}

The identity
\begin{equation}
  \frac{\mathrm{erfc}(\kappa r)}{r}
  =\frac{2}{\sqrt{\pi}}\int_\kappa^\infty e^{-t^2r^2}\,dt
\end{equation}
converts each real-space Ewald term into an auxiliary Gaussian in $\vrr$,
at the cost of a one-dimensional $t$-integral.

\subsection{Augmented nonlinear parameter matrix}

The factor $e^{-t^2|\vP_{ij}^T\vrr+\vn\cdot\vL|^2}$ is a Gaussian in $\vrr$
with rank-1 increment $t^2\vP_{ij}\vP_{ij}^T$.  The combined
matrix is $\At_{kl}^{(t,ij)}=[A_{kl}^{(t,ij)}]\otimes I_3$ with
$A_{kl}^{(t,ij)}$ given by Eq.~\eqref{eq:Atij}.

\paragraph{Matrix determinant lemma.}
With $\mathbf{w}=\ve_i-\ve_j$, the rank-1 determinant formula gives
Eq.~\eqref{eq:detlem}.

\paragraph{Sherman--Morrison formula.}
\begin{equation}
  (A_{kl}^{(t,ij)})^{-1}=A_{kl}^{-1}
  -\frac{t^2\,A_{kl}^{-1}\mathbf{w}\mathbf{w}^TA_{kl}^{-1}}{
  1+t^2\sigma_{ij,s}^2}.
\end{equation}

\subsection{Completing the square and quadratic form}

After completing the square the combined exponent evaluates to
$-\mathcal{Q}_{\vM}^{(t,ij,\vn)}$ as given in Eq.~\eqref{eq:Qee}, and the
$3n$-dimensional Gaussian integral yields the prefactor
$\mathcal{S}_{kl}/(1+t^2\sigma_{ij,s}^2)^{3/2}$, leading to
Eq.~\eqref{eq:Vee_real}.  The electron–nuclear result Eq.~\eqref{eq:VeN_real}
follows from the same derivation with the substitutions \eqref{eq:eN_replace}.

\section{Derivation of the Neutral Coulomb Matrix Elements}
\label{app:neutral_coulomb}

\subsection{Setup and applicability}

For a charge-neutral simulation cell the bare periodic Coulomb sum converges
absolutely and no Ewald decomposition is required.  By Corollary~1a of
Theorem~1, the matrix element of the operator $\hat{V}^{(ee)}=\sum_{i=1}^{n-1}\sum_{j=i+1}^{n}|\vrr_i-\vrr_j|^{-1}$
(lattice-periodic by overall charge neutrality) satisfies Eq.~\eqref{eq:unfolding},
giving, for each pair $(i,j)$,
\begin{eqnarray}\label{eq:J_def}
  V_{kl}^{(ee,\mathrm{bare})}& =& \sum_{i=1}^{n-1}\sum_{j=i+1}^{n}\sum_\vM
  \mathcal{J}_\vM^{(ij)}, \quad
  \mathcal{J}_\vM^{(ij)}\nonumber \\&=&\int_{\mathbb{R}^{3n}}
  \frac{\phi_k(\vrr)\,\phi_l(\vrr-\vT_\vM)}{|\vP_{ij}^T\vrr|}\,d\vrr.
\end{eqnarray}

\subsection{Integral representation of $1/r$ and evaluation}

Using $1/r=(2/\sqrt{\pi})\int_0^\infty e^{-t^2r^2}dt$ and the
$\vn=\mathbf{0}$ term of the Ewald appendix (Appendix~\ref{app:real_coulomb}):
\begin{equation}\label{eq:J_tint}
  \mathcal{J}_\vM^{(ij)}
  = \frac{2}{\sqrt{\pi}}\,\omega_\vM\,\mathcal{S}_{kl}
    \int_0^\infty
    \frac{\exp\!\bigl[-t^2|\bar{u}_{ij,\vM}|^2/(1+t^2\sigma_{ij,s}^2)\bigr]}
         {(1+t^2\sigma_{ij,s}^2)^{3/2}}\,dt.
\end{equation}
The substitution $s = t\sigma_{ij,s}/\sqrt{1+t^2\sigma_{ij,s}^2}$ maps
$t\in[0,\infty)$ to $s\in[0,1)$ and gives $(1+t^2\sigma^2)^{-3/2}dt
= (1/\sigma)ds$, so
\begin{eqnarray}
  \frac{2}{\sqrt{\pi}}\int_0^\infty\!\!
  \frac{e^{-t^2R^2/(1+t^2\sigma^2)}}{(1+t^2\sigma^2)^{3/2}}\,dt
  &=& \frac{2}{\sqrt{\pi}\sigma}\int_0^1 e^{-s^2R^2/\sigma^2}\,ds
  \nonumber \\
  &=& \frac{\mathrm{erf}(R/\sigma)}{R},
\end{eqnarray}
where $R=|\bar{u}_{ij,\vM}|$ and $\sigma=\sigma_{ij,s}$.  Hence
\begin{equation}\label{eq:J_final}
  \mathcal{J}_\vM^{(ij)}
  = \omega_\vM\,\mathcal{S}_{kl}\,
    \frac{\mathrm{erf}(|\bar{u}_{ij,\vM}|/\sigma_{ij,s})}{|\bar{u}_{ij,\vM}|},
\end{equation}
with $F_0(x)=\frac{\sqrt{\pi}}{2}\mathrm{erf}(\sqrt{x})/\sqrt{x}$ giving
the equivalent compact form $\mathcal{J}=(2/\sigma_{ij,s})\omega_\vM\mathcal{S}_{kl}
F_0(R^2/\sigma^2)$.

The finite limit $\lim_{R\to 0}\mathrm{erf}(R/\sigma)/R = 2/(\sigma\sqrt{\pi})$
confirms regularity at zero pair separation.  The electron–nuclear case
follows identically with $\sigma_{ij,s}\to\sigma_i$,
$|\bar{u}_{ij,\vM}|\to|\bar{\vrr}_{\vM,i}-\vRR_I|$, giving
Eqs.~\eqref{eq:Vee_bare}--\eqref{eq:VeN_bare}.

\subsection{Connection to the Ewald real-space formula}

Setting $\kappa=0$ and $\vn=\mathbf{0}$ in Eq.~\eqref{eq:Vee_real} gives
$\mathrm{erfc}(0)/R=1/R$, recovering the full $[0,\infty)$ integral above.
For a neutral cell all $\vn\neq\mathbf{0}$ shells and the entire
reciprocal-space sum vanish as $\kappa\to 0$, so the Ewald formula reduces
exactly to Eqs.~\eqref{eq:Vee_bare}--\eqref{eq:VeN_bare}, confirming
consistency.

\section{Delta-Function Matrix Elements and the Coulomb Potential}
\label{app:delta}

\subsection{Overview}

The contact operators $\delta^{(3)}(\vrr_i-\vrr_j)$ and
$\delta^{(3)}(\vrr_i-\mathbf{S})$ have closed-form matrix elements in the
SCG basis.  More importantly, the Coulomb operator $1/|\vrr_i-\vrr_j|$ can
be expressed as a weighted integral of $\delta^{(3)}(\vrr_i-\vrr_j-\mathbf{u})$
over $\mathbf{u}$, providing a third independent derivation of the neutral
Coulomb result and establishing exact equivalence with
Appendix~\ref{app:neutral_coulomb}.

\subsection{Matrix element of $\delta^{(3)}(\vrr_i - \vrr_j)$}

After unfolding (Theorem~1, Corollary~1a), the per-image integral is
\begin{equation}
  \mathcal{D}_\vM^{(ij)}
  = \int_{\mathbb{R}^{3n}}\phi_k(\vrr)\,
    \delta^{(3)}(\vrr_i-\vrr_j)\,\phi_l(\vrr-\vT_\vM)\,d\vrr.
\end{equation}
Using the combined Gaussian product and the Fourier representation
$\delta^{(3)}(\mathbf{u})=(2\pi)^{-3}\int e^{i\mathbf{q}\cdot\mathbf{u}}d^3\mathbf{q}$
with $\mathbf{u}=\vP_{ij}^T\vrr$:
\begin{widetext}
\begin{align}
  \mathcal{D}_\vM^{(ij)}
  &= \omega_\vM\int\!\frac{d^3\mathbf{q}}{(2\pi)^3}\,
     e^{i\mathbf{q}\cdot\bar{u}_{ij,\vM}}
     \int_{\mathbb{R}^{3n}}
     e^{-(\vrr-\bar\vrr_\vM)^T\At_{kl}(\vrr-\bar\vrr_\vM)+i\mathbf{q}\cdot\vP_{ij}^T(\vrr-\bar\vrr_\vM)}
     d\vrr.
\end{align}
\end{widetext}
The inner Gaussian integral with linear phase $i(\vP_{ij}\mathbf{q})^T(\vrr-\bar\vrr_\vM)$
gives $\mathcal{S}_{kl}\exp(-\mathbf{q}^T\vP_{ij}^T\At_{kl}^{-1}\vP_{ij}\mathbf{q}/4)
=\mathcal{S}_{kl}e^{-\sigma_{ij,s}^2|\mathbf{q}|^2/4}$.  The remaining
$\mathbf{q}$-integral is an inverse Fourier transform of a Gaussian:
\begin{equation}\label{eq:delta_ij_result}
  \boxed{\mathcal{D}_\vM^{(ij)}
  = \omega_\vM\,\mathcal{S}_{kl}\,
    \frac{1}{(\pi\sigma_{ij,s}^2)^{3/2}}\,
    \exp\!\left(-\frac{|\bar{u}_{ij,\vM}|^2}{\sigma_{ij,s}^2}\right).}
\end{equation}
The full matrix element is
\begin{equation}\label{eq:Dkl_ij}
  \langle\Phi_k|\delta^{(3)}(\vrr_i-\vrr_j)|\Phi_l\rangle
  = \frac{\mathcal{S}_{kl}}{(\pi\sigma_{ij,s}^2)^{3/2}}
    \sum_\vM\omega_\vM\,
    e^{-|\bar{u}_{ij,\vM}|^2/\sigma_{ij,s}^2}.
\end{equation}
The result is the combined Gaussian weight $\omega_\vM\mathcal{S}_{kl}$
times a three-dimensional Gaussian in the mean pair displacement
$\bar{u}_{ij,\vM}$ with width $\sigma_{ij,s}/\sqrt{2}$.

\subsection{Matrix element of $\delta^{(3)}(\vrr_i - \mathbf{S})$}

For the operator $\delta^{(3)}(\vrr_i-\mathbf{S})$ with a fixed vector
$\mathbf{S}\in\mathbb{R}^3$, the projector is $\vP_i=\ve_i\otimes I_3$,
$\vP_i^T\At_{kl}^{-1}\vP_i=\sigma_i^2 I_3$, and the mean is
$\bar{\vrr}_{\vM,i}$.  The identical Fourier argument gives
\begin{eqnarray}\label{eq:delta_iS_result}
 \mathcal{D}_\vM^{(i,\mathbf{S})}
  &\equiv&\int_{\mathbb{R}^{3n}}\phi_k(\vrr)\,
    \delta^{(3)}(\vrr_i-\mathbf{S})\,\phi_l(\vrr-\vT_\vM)\,d\vrr
  \nonumber\\ &=& \omega_\vM\,\mathcal{S}_{kl}\,
    \frac{e^{-|\bar{\vrr}_{\vM,i}-\mathbf{S}|^2/\sigma_i^2}}
         {(\pi\sigma_i^2)^{3/2}},
\end{eqnarray}
and
\begin{equation}\label{eq:Dkl_iS}
  \langle\Phi_k|\delta^{(3)}(\vrr_i-\mathbf{S})|\Phi_l\rangle
  = \frac{\mathcal{S}_{kl}}{(\pi\sigma_i^2)^{3/2}}
    \sum_\vM\omega_\vM\,
    e^{-|\bar{\vrr}_{\vM,i}-\mathbf{S}|^2/\sigma_i^2}.
\end{equation}
This is a Gaussian in $\mathbf{S}$ centred at $\bar{\vrr}_{\vM,i}$,
reflecting the probability density for finding electron $i$ at position
$\mathbf{S}$ in the combined Gaussian state.

\subsection{Coulomb potential from the delta-function matrix element}

The $1/r$ operator admits the resolution
\begin{equation}\label{eq:coulomb_from_delta}
  \frac{1}{|\vrr_i-\vrr_j|}
  = \int_{\mathbb{R}^3}\frac{\delta^{(3)}(\vrr_i-\vrr_j-\mathbf{u})}{|\mathbf{u}|}\,d^3\mathbf{u}.
\end{equation}
Inserting this into the matrix element and exchanging the order of integration:
\begin{equation}
  \mathcal{J}_\vM^{(ij)}
  = \int_{\mathbb{R}^3}\frac{\mathcal{D}_\vM^{(ij,\mathbf{u})}}{|\mathbf{u}|}\,d^3\mathbf{u},
\end{equation}
where $\mathcal{D}_\vM^{(ij,\mathbf{u})}$ denotes the matrix element of
$\delta^{(3)}(\vrr_i-\vrr_j-\mathbf{u})$.  By Eq.~\eqref{eq:delta_ij_result}
with $\bar{u}_{ij,\vM}\to\bar{u}_{ij,\vM}-\mathbf{u}$:
\begin{equation}
  \mathcal{D}_\vM^{(ij,\mathbf{u})}
  = \frac{\omega_\vM\,\mathcal{S}_{kl}}{(\pi\sigma_{ij,s}^2)^{3/2}}
    \exp\!\left(-\frac{|\bar{u}_{ij,\vM}-\mathbf{u}|^2}{\sigma_{ij,s}^2}\right).
\end{equation}
Hence
\begin{equation}\label{eq:J_via_delta}
  \mathcal{J}_\vM^{(ij)}
  = \frac{\omega_\vM\,\mathcal{S}_{kl}}{(\pi\sigma_{ij,s}^2)^{3/2}}
    \int_{\mathbb{R}^3}
    \frac{e^{-|\bar{u}_{ij,\vM}-\mathbf{u}|^2/\sigma_{ij,s}^2}}{|\mathbf{u}|}\,d^3\mathbf{u}.
\end{equation}

\subsection{Evaluation and equivalence}

The remaining integral is of the standard form
$\int_{\mathbb{R}^3}e^{-|\mathbf{c}-\mathbf{u}|^2/a^2}/|\mathbf{u}|\,d^3\mathbf{u}$
with $\mathbf{c}=\bar{u}_{ij,\vM}$ and $a=\sigma_{ij,s}$.  Substituting
$\mathbf{u}=a\mathbf{t}$ and using the standard result
\begin{equation}\label{eq:gauss_coulomb_conv}
  \int_{\mathbb{R}^3}\frac{e^{-|\mathbf{t}-\mathbf{R}|^2}}{|\mathbf{t}|}\,d^3\mathbf{t}
  = \frac{\pi^{3/2}\,\mathrm{erf}(|\mathbf{R}|)}{|\mathbf{R}|},
\end{equation}
with $\mathbf{R}=\mathbf{c}/a=\bar{u}_{ij,\vM}/\sigma_{ij,s}$:
\begin{eqnarray}
  \int_{\mathbb{R}^3}
  \frac{e^{-|\mathbf{c}-\mathbf{u}|^2/a^2}}{|\mathbf{u}|}\,d^3\mathbf{u}\\
  &=& a^3\cdot\frac{1}{a}\int\frac{e^{-|\mathbf{t}-\mathbf{R}|^2}}{|\mathbf{t}|}\,d^3\mathbf{t}
  \notag\\
  \nonumber \\ &=& \frac{a^2\pi^{3/2}\,\mathrm{erf}(|\mathbf{c}|/a)}{|\mathbf{c}|/a}
  \nonumber \\ &=& \frac{\pi^{3/2}\sigma_{ij,s}^3\,\mathrm{erf}(R/\sigma_{ij,s})}{R},
\end{eqnarray}
where $R=|\bar{u}_{ij,\vM}|$.  Substituting into Eq.~\eqref{eq:J_via_delta}:
\begin{eqnarray}
  \mathcal{J}_\vM^{(ij)}
  &=& \frac{\omega_\vM\,\mathcal{S}_{kl}}{(\pi\sigma_{ij,s}^2)^{3/2}}
    \cdot
    \frac{\pi^{3/2}\sigma_{ij,s}^3\,\mathrm{erf}(R/\sigma_{ij,s})}{R}
  \nonumber \\ &=& \omega_\vM\,\mathcal{S}_{kl}\,\frac{\mathrm{erf}(R/\sigma_{ij,s})}{R}.
\end{eqnarray}
This is identical to Eq.~\eqref{eq:J_final}, confirming exact equivalence
between the $t$-integral method (Appendix~\ref{app:neutral_coulomb}) and
the delta-function convolution method. \hfill$\square$

\paragraph{Proof of the key convolution identity \eqref{eq:gauss_coulomb_conv}.}
Write $1/|\mathbf{t}|=(2/\sqrt{\pi})\int_0^\infty e^{-\lambda^2|\mathbf{t}|^2}d\lambda$
and evaluate the Gaussian integral over $\mathbf{t}$:
$\int e^{-|\mathbf{t}-\mathbf{R}|^2-\lambda^2|\mathbf{t}|^2}d^3\mathbf{t}
=[\pi/(1+\lambda^2)]^{3/2}e^{-\lambda^2|\mathbf{R}|^2/(1+\lambda^2)}$.
Then
$\frac{2}{\sqrt{\pi}}\int_0^\infty[\pi/(1+\lambda^2)]^{3/2}e^{-\lambda^2R^2/(1+\lambda^2)}d\lambda
= \pi^{3/2}\mathrm{erf}(R)/R$
by the same substitution $s=\lambda/\sqrt{1+\lambda^2}$ used in
Appendix~\ref{app:neutral_coulomb}. \hfill$\square$

\subsection{Electron--nuclear Coulomb via $\delta(\vrr_i-\mathbf{S})$}

The electron--nuclear potential $1/|\vrr_i-\vRR_I|$ can similarly be written as
\begin{equation}
  \frac{1}{|\vrr_i-\vRR_I|}
  = \int_{\mathbb{R}^3}\frac{\delta^{(3)}(\vrr_i-\mathbf{S})}{|\mathbf{S}-\vRR_I|}\,d^3\mathbf{S}.
\end{equation}
Using Eq.~\eqref{eq:delta_iS_result}:
\begin{equation}
  \mathcal{J}_\vM^{(i,I)}
  = \frac{\omega_\vM\,\mathcal{S}_{kl}}{(\pi\sigma_i^2)^{3/2}}
    \int_{\mathbb{R}^3}
    \frac{e^{-|\bar{\vrr}_{\vM,i}-\mathbf{S}|^2/\sigma_i^2}}{|\mathbf{S}-\vRR_I|}\,d^3\mathbf{S}.
\end{equation}
Applying identity~\eqref{eq:gauss_coulomb_conv} with $\mathbf{R}=(\bar{\vrr}_{\vM,i}-\vRR_I)/\sigma_i$
and $\mathbf{S}-\vRR_I\to\sigma_i\mathbf{t}$, $\bar{\vrr}_{\vM,i}-\mathbf{S}=-\sigma_i(\mathbf{t}-\mathbf{R})$:
\begin{equation}
  \mathcal{J}_\vM^{(i,I)}
  = \omega_\vM\,\mathcal{S}_{kl}\,
    \frac{\mathrm{erf}(|\bar{\vrr}_{\vM,i}-\vRR_I|/\sigma_i)}{|\bar{\vrr}_{\vM,i}-\vRR_I|},
\end{equation}
recovering Eq.~\eqref{eq:VeN_bare} upon multiplication by $-Z_I$ and
summation over $i$, $I$, $\vM$. \hfill$\square$

\subsection{Structural interpretation}

Equations~\eqref{eq:delta_ij_result} and \eqref{eq:delta_iS_result} reveal
the following structure.  The SCG basis smears each electron-pair contact
over a 3D Gaussian of width $\sigma_{ij,s}/\sqrt{2}$.  The Coulomb matrix
element is the convolution of this smeared contact density with the bare
$1/r$ kernel, giving the screened potential $\mathrm{erf}(R/\sigma)/R$.  The
effective range $\sigma_{ij,s}=\sqrt{(A_{kl}^{-1})_{ii}+(A_{kl}^{-1})_{jj}-2(A_{kl}^{-1})_{ij}}$
is the standard deviation of the pair coordinate $\vrr_i-\vrr_j$ in the
combined Gaussian; a more correlated basis (larger off-diagonal $A_k$)
produces a narrower smearing and a Coulomb matrix element closer to the
bare $1/r$ result.

\section{Gaussian Integral Identities}
\label{app:gaussians}

For $M\succ 0$ (positive definite, $N$-dimensional):
\begin{align}
  \int_{\mathbb{R}^N}e^{-\vx^TM\vx}\,d\vx
    &=\frac{\pi^{N/2}}{(\det M)^{1/2}},\\
  \int_{\mathbb{R}^N}e^{-\vx^TM\vx+\mathbf{b}^T\vx}\,d\vx
    &=\frac{\pi^{N/2}}{(\det M)^{1/2}}\,e^{\mathbf{b}^TM^{-1}\mathbf{b}/4},\\
  \int_{\mathbb{R}^N}x_\alpha x_\beta\,e^{-\vx^TM\vx}\,d\vx
    &=\frac{\pi^{N/2}}{(\det M)^{1/2}}\,\frac{1}{2}(M^{-1})_{\alpha\beta}.
\end{align}

The product of two Gaussians centered at $\mathbf{a}$ and $\mathbf{b}$:
\begin{multline}
  e^{-(\vx-\mathbf{a})^TM_1(\vx-\mathbf{a})-(\vx-\mathbf{b})^TM_2(\vx-\mathbf{b})}\\
  =e^{-(\mathbf{a}-\mathbf{b})^TM_1(M_1+M_2)^{-1}M_2(\mathbf{a}-\mathbf{b})}
   \cdot e^{-(\vx-\bar{\vx})^T(M_1+M_2)(\vx-\bar{\vx})},
\end{multline}
with combined center $\bar{\vx}=(M_1+M_2)^{-1}(M_1\mathbf{a}+M_2\mathbf{b})$.

\section{Gradient Formulas for Basis Optimization}
\label{app:gradients}

The variational parameters $\{A_k,\vs_k\}$ are optimized by minimizing
$E_0=\mathbf{c}^T\mathbf{H}\mathbf{c}/(\mathbf{c}^T\mathbf{S}\mathbf{c})$.
The gradient of $E_0$ with respect to any parameter $\theta$ is
\begin{equation}
  \frac{\partial E_0}{\partial\theta}
  =\mathbf{c}^T\!\Bigl(\frac{\partial\mathbf{H}}{\partial\theta}
                       -E_0\frac{\partial\mathbf{S}}{\partial\theta}\Bigr)\mathbf{c}.
\end{equation}

\paragraph{Perturbations of composite quantities.}
Under a variation $\delta A_k$ (induced by $\delta L_k$ via $\delta A_k=\delta L_k L_k^T+L_k\delta L_k^T$):
\begin{align}
  \delta A_{kl}^{-1} &= -A_{kl}^{-1}(\delta A_k)A_{kl}^{-1},\label{eq:dAklinv}\\
  \delta C_{kl} &= A_lA_{kl}^{-1}(\delta A_k)A_{kl}^{-1}A_l,\\
  \frac{\delta\det A_{kl}}{\det A_{kl}} &= \Tr(A_{kl}^{-1}\delta A_k).\label{eq:ddet}
\end{align}

\paragraph{Derivative of overlap.}
\begin{align}
  \frac{\partial S_{kl}}{\partial(\vs_k)_\alpha}
  &= -2\,\mathcal{S}_{kl}\sum_{\vM}\omega_{\vM}(\widetilde{C}_{kl}\vd_{\vM})_\alpha,\\
  \frac{\partial S_{kl}}{\partial(A_k)_{pq}}
  &= S_{kl}\Bigl[-\frac{3}{2}(A_{kl}^{-1})_{pq}\Bigr]
     -\mathcal{S}_{kl}\sum_{\vM}\omega_{\vM}\,
      \vd_{\vM}^T\frac{\partial\widetilde{C}_{kl}}{\partial(A_k)_{pq}}\vd_{\vM}.
\end{align}
Higher derivatives and the kinetic-energy gradient follow from the product
rule applied to Eq.~\eqref{eq:Tkl}, using Eqs.~\eqref{eq:dAklinv}–\eqref{eq:ddet}.

\section{Convergence Acceleration: Theta-Function Dual}
\label{app:convergence}
\subsection{The problem: slow convergence for diffuse basis functions}

Every matrix element derived in the paper reduces, after unfolding
(Theorem~1), to a lattice sum of the form
\begin{equation}\label{eq:imagesum}
  \mathcal{I}
  = \sum_{M\in\mathbb{Z}^{3n}} \omega_M\,f(d_M),
  \qquad
  \omega_M = e^{-d_M^{\top}\tilde{C}_{kl}\,d_M},
  \quad
  d_M = s_k - s_l - T_M,
\end{equation}
where $\tilde{C}_{kl} = C_{kl}\otimes I_3$ with
$C_{kl}=A_k A_{kl}^{-1}A_l$ positive definite, and $f(d_M)$ is an
operator-specific function (a polynomial for kinetic energy, a screened
Coulomb factor for potential energy, etc.).

Convergence of~\eqref{eq:imagesum} is governed by the smallest
eigenvalue $c_{\min}$ of $C_{kl}$: the number of image shells needed
scales as
\begin{equation}
  N_{\mathrm{shells}} \sim
  \left(\frac{\chi_{\mathrm{cut}}}{\sqrt{c_{\min}}}\right)^{3n},
\end{equation}
where $\chi^2_{\mathrm{cut}}\approx 20$--$30$ is the truncation
threshold for double precision.  When basis functions are
\emph{diffuse} ($A_k$ has small eigenvalues), $C_{kl}$ inherits small
eigenvalues, $c_{\min}\to 0$, and $N_{\mathrm{shells}}$ diverges.
This is the regime where the direct-space sum becomes prohibitively
expensive and a dual representation is needed.

\subsection{The diagonal case: Jacobi theta functions}

\subsubsection{Factorization of the sum}

When $C_{kl}$ is diagonal,
$C_{kl}=\mathrm{diag}(c_1,\dots,c_n)$, and the shift difference is
$d_M = (s_{k,1}-s_{l,1}-m_1 L,\dots)$, the image weight factorizes
over the $3n$ Cartesian directions $\alpha=1,\dots,3n$:
\begin{equation}\label{eq:factored}
  \sum_{M\in\mathbb{Z}^{3n}} \omega_M
  = \prod_{\alpha=1}^{3n}
    \underbrace{\sum_{m_\alpha\in\mathbb{Z}}
    e^{-c_\alpha(s_{k,\alpha}-s_{l,\alpha}-m_\alpha L_\alpha)^2}}_{
    \displaystyle =:\,\vartheta_\alpha}.
\end{equation}
Each factor $\vartheta_\alpha$ is a \emph{Jacobi theta function}.
Recall the standard definition
\begin{equation}\label{eq:theta3}
  \vartheta_3(z\,|\,\tau)
  = \sum_{m\in\mathbb{Z}} e^{i\pi\tau m^2 + 2\pi i m z}.
\end{equation}
Setting
\begin{equation}
  \tau_\alpha = \frac{ic_\alpha L_\alpha^2}{\pi},
  \qquad
  z_\alpha    = \frac{ic_\alpha(s_{k,\alpha}-s_{l,\alpha})L_\alpha}{\pi},
\end{equation}
each factor in~\eqref{eq:factored} equals
$\vartheta_3(z_\alpha|\tau_\alpha)$ up to an overall exponential
prefactor, so
\begin{equation}\label{eq:product_theta}
  \sum_{M\in\mathbb{Z}^{3n}} \omega_M
  = \prod_{\alpha=1}^{3n} \vartheta_3(z_\alpha\,|\,\tau_\alpha).
\end{equation}

\subsubsection{The Jacobi imaginary transformation}

The theta function satisfies the \emph{Jacobi imaginary
transformation} (Eq.~J1 of the paper):
\begin{equation}\label{eq:jacobi}
  \vartheta_3(z\,|\,\tau)
  = \frac{1}{\sqrt{-i\tau}}\,
    e^{-iz^2/(\pi\tau)}\,
    \vartheta_3\!\left(\frac{z}{\tau}\,\bigg|\,-\frac{1}{\tau}\right).
\end{equation}

\noindent
This identity maps a theta function with modular parameter $\tau$ to
one with parameter $-1/\tau$.  The direct sum
$\vartheta_3(z|\tau)$ converges like $e^{-\pi|\mathrm{Im}\,\tau|m^2}$,
while the dual sum $\vartheta_3(z/\tau|-1/\tau)$ converges like
$e^{-\pi|\mathrm{Im}(-1/\tau)|m^2} = e^{-\pi m^2/|\mathrm{Im}\,\tau|}$.
Their convergence rates are therefore reciprocal.

\subsubsection{Convergence analysis}

Identifying $\mathrm{Im}\,\tau_\alpha = c_\alpha L_\alpha^2/\pi$:

\begin{center}
\renewcommand{\arraystretch}{1.6}
\begin{tabular}{lccc}
\toprule
Regime & $c_\alpha$ & Direct sum & Dual sum \\
\midrule
Diffuse basis & small & {slow:
  $e^{-c_\alpha L^2 m^2}$} & {fast:
  $e^{-\pi^2 m^2/(c_\alpha L^2)}$} \\
Compact basis & large & {fast:
  $e^{-c_\alpha L^2 m^2}$} & {slow:
  $e^{-\pi^2 m^2/(c_\alpha L^2)}$} \\
\bottomrule
\end{tabular}
\end{center}

\noindent
One always evaluates whichever representation converges faster.  The
crossover occurs when the two rates are equal, i.e.\
$c_\alpha L_\alpha^2 = \pi$, or equivalently
$\sqrt{c_\alpha} = \pi/L_\alpha$.

\subsection{The general case: Poisson summation formula}

When $C_{kl}$ is not diagonal (the generic situation for correlated
Gaussians with off-diagonal $A_k$), the sum~\eqref{eq:imagesum} does
not factorize.  The multidimensional generalization of the Jacobi
transformation is the \emph{Poisson summation formula} applied to the
Gaussian function $g(x) = e^{-x^\top \tilde{C}_{kl}\,x}$.

\subsubsection{Result}

\begin{equation}\label{eq:poisson}
  \sum_{M\in\mathbb{Z}^{3n}} e^{-d_M^{\top}\tilde{C}_{kl}\,d_M}
  = \frac{\pi^{3n/2}}{(\det C_{kl})^{3/2}\,\Omega^n}
    \sum_{K\in\mathcal{G}^n}
    e^{-K^{\top}(\tilde{C}_{kl})^{-1}K/4}\,
    e^{iK^{\top}(s_k-s_l)},
\end{equation}

\noindent
where:
\begin{itemize}
  \item $\mathcal{G}^n$ is the $3n$-dimensional reciprocal lattice
        with vectors
        \[
          K = 2\pi\!\left(\frac{k_{1,x}}{L_x},\frac{k_{1,y}}{L_y},
              \frac{k_{1,z}}{L_z},\dots,
              \frac{k_{n,x}}{L_x},\frac{k_{n,y}}{L_y},
              \frac{k_{n,z}}{L_z}\right),
          \quad k_{i,\mu}\in\mathbb{Z};
        \]
  \item $\Omega^n = (L_x L_y L_z)^n$ is the $n$-electron cell volume;
  \item $\det C_{kl}$ enters via the standard Gaussian Fourier
        transform identity
        $\int_{\mathbb{R}^N} e^{-x^\top C x}\,dx
         = \pi^{N/2}/(\det C)^{1/2}$.
\end{itemize}

\subsubsection{Derivation sketch}

The Poisson summation formula states that for any sufficiently regular
function $g:\mathbb{R}^N\to\mathbb{C}$ and lattice
$\Lambda\subset\mathbb{R}^N$ with dual lattice $\Lambda^*$:
\begin{equation}
  \sum_{v\in\Lambda} g(x+v)
  = \frac{1}{\mathrm{vol}(\Lambda)}
    \sum_{K\in\Lambda^*} \hat{g}(K)\,e^{iK\cdot x},
\end{equation}
where $\hat{g}(K) = \int_{\mathbb{R}^N} g(y)\,e^{-iK\cdot y}\,dy$ is
the Fourier transform.  Applying this with
$g(y)=e^{-y^\top \tilde{C}_{kl} y}$,
$\Lambda = \{T_M : M\in\mathbb{Z}^{3n}\}$,
$x = s_k-s_l$, and using
\begin{equation}
  \hat{g}(K)
  = \int_{\mathbb{R}^{3n}} e^{-y^\top \tilde{C}_{kl}y - iK\cdot y}\,dy
  = \frac{\pi^{3n/2}}{(\det C_{kl})^{3/2}}\,
    e^{-K^\top(\tilde{C}_{kl})^{-1}K/4}
\end{equation}
yields~\eqref{eq:poisson} directly.

\subsection{Convergence of the dual sum and optimal switching}

\subsubsection{Convergence rates}

Let $c_{\min}$ and $c_{\max}^{-1}$ denote the smallest eigenvalue of
$C_{kl}$ and the largest eigenvalue of $C_{kl}^{-1}$ respectively.
The number of significant terms in each representation is:
\begin{equation}
  N_{\mathrm{direct}} \sim
  \left(\frac{\chi_{\mathrm{cut}}}{\sqrt{c_{\min}}}\right)^{3n},
  \qquad
  N_{\mathrm{dual}} \sim
  \left(\frac{2\chi_{\mathrm{cut}}\sqrt{c_{\max}}}{2\pi/L}\right)^{3n},
\end{equation}
where $L$ is a representative cell dimension and $\chi^2_{\mathrm{cut}}
\approx 20$--$30$.  The product $N_{\mathrm{direct}}\cdot
N_{\mathrm{dual}}$ is independent of $c_{\min}$, confirming the
reciprocal nature of the two representations.

\subsubsection{Optimal switching criterion}

Equating $N_{\mathrm{direct}} = N_{\mathrm{dual}}$ gives the crossover
condition:
\begin{equation}\label{eq:switch}
  c_{\min}\,c_{\max} \approx \frac{\pi^2}{L^2}.
\end{equation}
In practice one evaluates
\begin{equation}
  \text{Use direct sum if } c_{\min} \gtrsim \frac{\pi}{L^2},
  \quad
  \text{use dual sum otherwise.}
\end{equation}

\subsection{Connection to the Ewald method}

The Poisson summation approach for the image sum is the exact analogue
of Ewald summation for the Coulomb lattice sum.  Table~\ref{tab:analogy}
summarizes the parallel.
\begin{widetext}
\begin{table}[h]
\centering
\renewcommand{\arraystretch}{1.6}
\begin{tabular}{lcc}
\toprule
& \textbf{Ewald (Coulomb sum)} & \textbf{Poisson (image sum)} \\
\midrule
Object being summed
  & $1/|\mathbf{u}+\mathbf{n}\cdot\mathbf{L}|$
  & $e^{-d_M^\top \tilde{C}_{kl}d_M}$ \\
Splitting device
  & $1 = \mathrm{erfc}(\kappa r)+\mathrm{erf}(\kappa r)$
  & Fourier transform of Gaussian \\
Real-space convergence
  & $e^{-\kappa^2|\mathbf{n}\cdot\mathbf{L}|^2}$
  & $e^{-c_{\min}|T_M|^2}$ \\
Reciprocal convergence
  & $e^{-G^2/4\kappa^2}$
  & $e^{-|K|^2/(4c_{\max})}$ \\
Free parameter
  & $\kappa$ (Ewald splitting)
  & switching threshold~\eqref{eq:switch} \\
Auxiliary integral needed?
  & yes ($\mathrm{erfc}/r$ via $t$-integral)
  & no (Gaussian FT is analytic) \\
\bottomrule
\end{tabular}
\caption{Analogy between Ewald summation and Poisson-dual image-sum
acceleration.}
\label{tab:analogy}
\end{table}
\end{widetext}
\noindent
The key advantage of the Poisson route for Gaussian basis sets is that
the Fourier transform of a Gaussian is again a Gaussian — no auxiliary
splitting parameter $\kappa$ or $t$-integration is required.  The dual
sum~\eqref{eq:poisson} is therefore algebraically simpler than the
Ewald decomposition, at the cost of requiring $C_{kl}$ to be
invertible (which is always guaranteed by positive definiteness).

\section{Connection to Pair-Correlation--Single-Particle Gaussian Bases}
\label{app:connection}

\subsection{The alternative basis form}

A second class of many-electron Gaussian basis functions appearing in the
literature takes the form
\begin{equation}\label{eq:psi_alt}
  \psi(\vrr) = \exp\!\left[
    -\tfrac{1}{2}\sum_{i,j=1}^{n} A_{ij}\,\vrr_i\cdot\vrr_j
    -\sum_{i=1}^{n} \beta_i\,|\vrr_i - \mathbf{s}_i|^2
  \right],
\end{equation}
where $A=(A_{ij})\in\mathbb{R}^{n\times n}$ is a symmetric matrix of
pair-coupling constants (not necessarily positive definite), each $\beta_i>0$
is a single-particle Gaussian width, and $\mathbf{s}_i\in\mathbb{R}^3$ is
the center of the $i$-th single-particle factor.  The first sum encodes
inter-electronic correlations through the unshifted quadratic
$\mathbf{r}_i\cdot\mathbf{r}_j$; the second sum is a product of Gaussians
each centered at $\mathbf{s}_i$.

We show that Eq.~\eqref{eq:psi_alt} is a special case of the SCG basis
Eq.~\eqref{eq:scg} and derive the exact correspondence between the two
parameter sets.

\subsection{Reduction to standard quadratic form}

Expand the single-particle terms:
$\beta_i|\vrr_i-\mathbf{s}_i|^2
=\beta_i|\vrr_i|^2 - 2\beta_i\vrr_i\cdot\mathbf{s}_i + \beta_i|\mathbf{s}_i|^2$.
Absorbing the constant $\sum_i\beta_i|\mathbf{s}_i|^2$ into an overall
normalization, the exponent of $\psi$ becomes
\begin{equation}
  -\tfrac{1}{2}\sum_{i,j}A_{ij}\,\vrr_i\cdot\vrr_j
  -\sum_i\beta_i|\vrr_i|^2
  +2\sum_i\beta_i\,\vrr_i\cdot\mathbf{s}_i.
\end{equation}
Combining the quadratic parts and introducing the diagonal matrix
$D_\beta=\mathrm{diag}(\beta_1,\dots,\beta_n)$:
\begin{equation}\label{eq:psi_exponent}
  -\tfrac{1}{2}\sum_{i,j}(A_{ij}+2\beta_i\delta_{ij})\,\vrr_i\cdot\vrr_j
  +\sum_i\,2\beta_i\,\vrr_i\cdot\mathbf{s}_i.
\end{equation}
Define the \emph{effective matrix}
\begin{equation}\label{eq:Beff}
  B \equiv \tfrac{1}{2}(A + 2D_\beta) \in\mathbb{R}^{n\times n}.
\end{equation}
Positive definiteness of $B$ (required for normalizability) demands that
$A+2D_\beta\succ 0$; equivalently, all eigenvalues of $A$ must exceed
$-2\min_i\beta_i$.  The exponent \eqref{eq:psi_exponent} is then
\begin{equation}
  -\vrr^T(B\otimes I_3)\vrr + 2\bigl(D_\beta\otimes I_3\bigr)\vs\cdot\vrr,
\end{equation}
where $\vs=(\mathbf{s}_1,\dots,\mathbf{s}_n)^T\in\mathbb{R}^{3n}$.  Completing
the square about the center
\begin{equation}\label{eq:psi_center}
  \bar{\vrr} = (B\otimes I_3)^{-1}(D_\beta\otimes I_3)\,\vs
            = (B^{-1}D_\beta\otimes I_3)\,\vs,
\end{equation}
the exponent becomes
\begin{equation}
  -(\vrr-\bar{\vrr})^T(B\otimes I_3)(\vrr-\bar{\vrr})
  + \bar{\vrr}^T(B\otimes I_3)\bar{\vrr}.
\end{equation}
Hence, up to the constant $e^{\bar{\vrr}^T(B\otimes I_3)\bar{\vrr}
-\sum_i\beta_i|\mathbf{s}_i|^2}$ (which is absorbed into the variational
coefficient),
\begin{equation}\label{eq:psi_as_phi}
  \psi(\vrr) \propto
  \exp\!\bigl[-(\vrr-\bar{\vrr})^T(B\otimes I_3)(\vrr-\bar{\vrr})\bigr].
\end{equation}
This is precisely the SCG form \eqref{eq:scg} with identification
\begin{equation}\label{eq:id_BtoA}
  \boxed{A_k = B = \tfrac{1}{2}(A+2D_\beta), \qquad
  \vs_k = \bar{\vrr} = (B^{-1}D_\beta\otimes I_3)\,\vs.}
\end{equation}

\subsection{Inverse map: SCG parameters to the alternative form}

Given an SCG with parameters $(A_k, \vs_k)$, the alternative-form parameters
$(A, D_\beta, \mathbf{s})$ satisfying Eq.~\eqref{eq:id_BtoA} are not unique:
$D_\beta$ is a free positive-definite diagonal matrix subject only to
$A=2A_k-2D_\beta$ being the desired pair-coupling matrix (which may have any
sign on its diagonal).  Choosing $D_\beta$ fixes everything:
\begin{equation}\label{eq:id_AtoB}
  A = 2(A_k - D_\beta), \qquad
  \vs_i = \frac{1}{\beta_i}\sum_{j=1}^n (A_k)_{ij}\,\mathbf{s}_{k,j}.
\end{equation}
Here $\mathbf{s}_{k,j}\in\mathbb{R}^3$ is the 3D block of $\vs_k$
corresponding to electron $j$.  The second relation follows from
$D_\beta\vs = A_k\vs_k$ (i.e.\ $B\bar{\vrr}=D_\beta\vs$ with $\bar{\vrr}=\vs_k$).

Three natural choices of $D_\beta$ are physically illuminating.

\paragraph{Choice 1: $D_\beta = \mathrm{diag}(A_k)$.}
The diagonal of $A_k$ is absorbed entirely into the single-particle widths,
$\beta_i=(A_k)_{ii}$, and the pair-coupling matrix becomes
$A_{ij}=2(A_k)_{ij}(1-\delta_{ij})$ — purely off-diagonal.  This is the
most natural separation of single-particle and correlation content.
The single-particle centers are
\begin{equation}
  \mathbf{s}_i = \frac{1}{(A_k)_{ii}}\sum_j (A_k)_{ij}\,\mathbf{s}_{k,j}.
\end{equation}

\paragraph{Choice 2: $D_\beta = A_k$.}
Here $A=0$ (no explicit pair coupling) and $\beta_i=(A_k)_{ii}$... but
this requires $A_k$ to be diagonal for self-consistency (since $A=2A_k-2A_k=0$
regardless, while $\vs_i=\mathbf{s}_{k,i}$).  More precisely, this choice
sets $\vs_i=(A_k)_{ii}^{-1}\sum_j(A_k)_{ij}\mathbf{s}_{k,j}$, which
reduces to $\vs_i=\mathbf{s}_{k,i}$ only if $A_k$ is diagonal.  When $A_k$
is diagonal the two bases are identical, recovering independent-particle
Gaussians $\prod_i e^{-\beta_i|\vrr_i-\mathbf{s}_i|^2}$.

\paragraph{Choice 3: Equal widths $\beta_i=\beta$ (all electrons same width).}
This is the simplest variational ansatz; with $D_\beta=\beta I$ and
$A=2(A_k-\beta I)$:
\begin{equation}
  \vs_i = \frac{1}{\beta}\sum_j(A_k)_{ij}\,\mathbf{s}_{k,j}.
\end{equation}


%

\end{document}